\documentclass[prb,twocolumn,superscriptaddress,showpacs]{revtex4}
\usepackage{graphicx}
\usepackage{amsmath}
\usepackage{color}
\begin{document}

\title{Exciton condensation in strongly correlated electron bilayers}
\author{Louk Rademaker}
\email{rademaker@lorentz.leidenuniv.nl}
\affiliation{Institute-Lorentz for Theoretical Physics, Leiden University, PO Box 9506, Leiden, The Netherlands}
\author{Jeroen van den Brink}
\affiliation{Institute for Theoretical Solid State Physics, IFW Dresden, 01171 Dresden, Germany}
\affiliation{Department of Physics, TU Dresden, D-01062 Dresden, Germany}
\author{Jan Zaanen}
\affiliation{Institute-Lorentz for Theoretical Physics, Leiden University, PO Box 9506, Leiden, The Netherlands}
\author{Hans Hilgenkamp}
\affiliation{Institute-Lorentz for Theoretical Physics, Leiden University, PO Box 9506, Leiden, The Netherlands}
\affiliation{Faculty of Science and Technology and MESA+ Institute for Nanotechnology, University of Twente, P.O. Box 217, 7500 AE Enschede, The Netherlands}
\date{\today}

\begin{abstract}
We studied the possibility of exciton condensation in Mott insulating bilayers. In these strongly correlated systems an exciton is the bound state of a double occupied and empty site. In the strong coupling limit the exciton acts as a hard-core boson. Its physics are captured by the exciton $t-J$ model, containing an effective $XXZ$ model describing the exciton dynamics only. Using numerical simulations and analytical mean field theory we constructed the ground state phase diagram. Three homogeneous phases can be distinguished: the antiferromagnet, the exciton checkerboard crystal and the exciton superfluid. For most model parameters, however, we predict macroscopic phase separation between these phases. The exciton superfluid exists only for large exciton hopping energy. Additionally we studied the collective modes and susceptibilities of the three phases. In the superfluid phase we find the striking feature that the bandwidth of the spin-triplet excitations, potentially detectable by resonant inelastic x-ray scattering (RIXS), is proportional to the superfluid density. The superfluid phase mode is visible in the charge susceptibility, measurable by RIXS or electron energy loss spectroscopy (EELS).
\end{abstract}

\pacs{71.35.Lk, 71.27.+a, 73.20.Mf}

\maketitle

\section{Introduction}

Strongly correlated electron systems exhibit the highest attained superconducting transition temperatures currently known, and a rich variety of complex electronic phases\cite{Imada:1998p2790,Lee:2006p1688}. Many compounds among this family of Mott insulators, such as the cuprates, are quasi-two-dimensional layered materials. This renders them ideal candidates for bilayer exciton condensation, which is the topic of this publication.

The effort to achieve the condensation of excitons has a long history starting just after the discovery of BCS theory\cite{Blatt:1962p4000,Keldysh:1968p4790,Moskalenko:2000p4767}. An exciton is the bound state of an electron and a hole and as such it can Bose condense. The obvious advantage of considering excitons above Cooper pairs is the strong Coulomb attraction between the electron and the hole; allowing in principle for a much higher critical temperature. To reduce the exciton lifetime problems caused by electron-hole recombination, it has been suggested to spatially separate the electrons and holes in their own subsequent layers\cite{Shevchenko:1976p4950,Lozovik:1976p4951}. This indeed has resulted in the experimental realization of exciton condensates, first in the so-called quantum Hall bilayers\cite{Eisenstein:2004p4770} and more recently without an externally applied magnetic field in electrically gated, optically pumped semiconductor quantum wells\cite{High:2012p5349}.

The successes of exciton condensation in semiconductor 2DEG bilayer systems have led to many proposals for exciton condensation in alternative bilayer materials, such as gated topological insulators\cite{Seradjeh:2009p4980} or double layer graphene\cite{Lozovik:2008p4877,Zhang:2008p4895,Dillenschneider:2008p4896,Min:2008p4795,Kharitonov:2008p5044}. However, these proposals are limited to the BCS paradigm of weak coupling. 

On the other hand, Mott insulators provide a completely different route to exciton condensation\cite{Ribeiro2006,Pentcheva:2009p5025,Millis:2010p5231}. Naively one would expect that the localization of the electrons and holes leads to a higher critical temperature, since $T_c$ is determined by the competition between the electronic kinetic energy and the electron-hole attraction. But the physics of exciton condensation in Mott insulators is in fact much richer. 

\begin{figure}
	\includegraphics[width=\columnwidth]{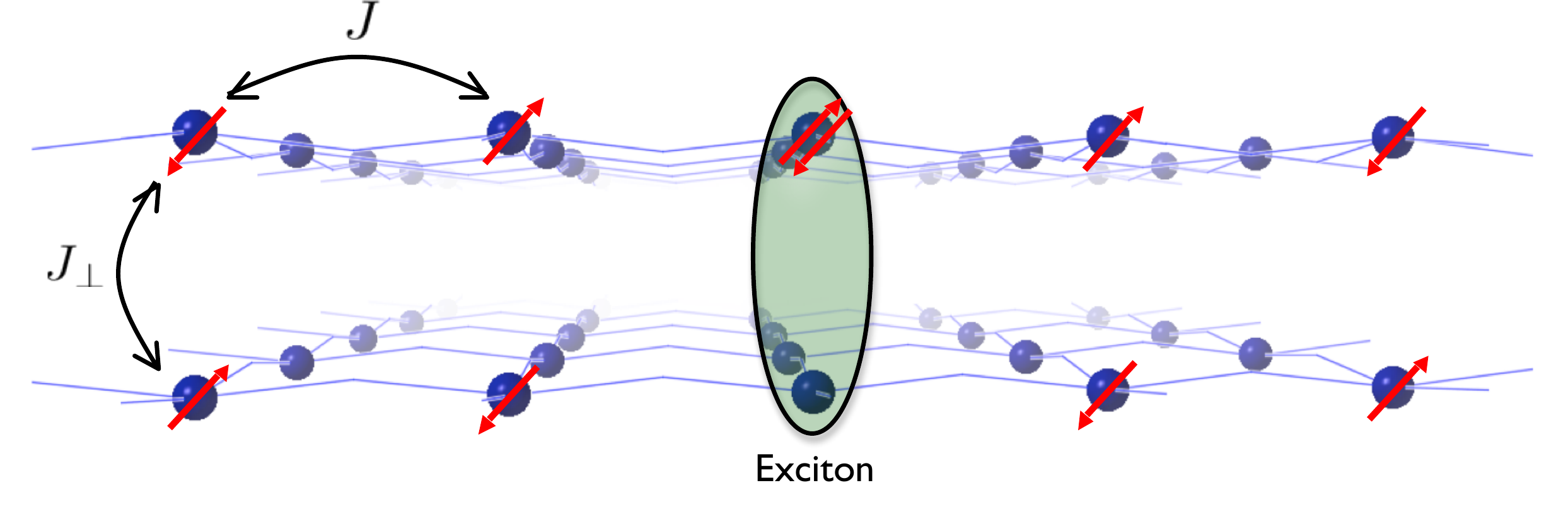}
	\caption{\label{Lattice}Side view of a strongly correlated electron bilayer with an exciton present. The red arrows denote the spin of the localized electrons, and the exciton is a bound state of a double occupied and an empty site.}
\end{figure}

Instead of the picture that the electron-hole pair lives in a conduction and valence band, an exciton now consists of a double occupied and vacant site bound together on an interlayer rung, see figure \ref{Lattice}. To estimate the binding energy, consider the in-plane charge-transfer excitons which are known to have a binding energy of the order of 1-2 eV\cite{ZhangNg}. Due to the small interlayer distances of order 1 nm we expect that a similar energy scale will set the binding of the interlayer exciton. As such, excitons in a Mott bilayer are most likely in the strongly coupled regime.

Furthermore, a single doublon-holon pair inserted into a Mott insulator leads to dynamical frustration effects\cite{Rademaker2012EPL, Rademaker2012arXiv}, even stronger than seen for a single hole in the $t-J$ model\cite{SingleHoleRefs,SingleHoleRefs2}. The study of excitons in strongly correlated materials thus catches the complexity of doped Mott insulators. As we discussed elsewhere \cite{Rademaker2012arXiv} the bosonic nature of the excitons actually falls short to completely eliminate all "fermion-like" signs: there are still left-over signs of the phase-string type\cite{Weng:2007p3423}. However, it is easy to demonstrate that collinear spin order is a sufficient condition for these signs to cancel out, leaving a truly bosonic dynamics controlling the ground state and long wavelength physics. The problem thereby reduces to that of hard-core bosons (the excitons) in a sign-free spin background.
This is very similar to the "spin-orbital" physics described by Kugel-Khomskii type models\cite{KugelKhomskii}, which can be viewed after all as describing d-d excitons 
interacting with spins. Also the lattice implementations\cite{VanDuinZaanen} of the SO(5) model\cite{ZhangSO5} for (cuprate) superconductivity are in this family.

Such bosonic problems can be handled with standard (semi-classical) mean field theory, and therefore the regime of finite exciton density can be addressed in an a-priori controlled manner. In most bilayer exciton set-ups, such as the quantum Hall bilayers or the pumped systems, there is no controllable equilibrium exciton density. In these cases one can hardly speak of the exciton density as a conserved quantity, and exciton condensation in the sense of spontaneously broken $U(1)$ symmetry is impossible\cite{Snoke:2006p4914}. However, in Mott insulators the dopant density per layer could be fixed by, for example, chemical doping. The effective exciton chemical potential is then by definition large compared to the recombination rate. Effectively, the excitons are at finite density in equilibrium and hence spontaneous $U(1)$ symmetry breaking is possible in the Mott insulating bilayer.

Besides the exciton superfluid phase one anticipates a plethora of competing orders, as is customary in strongly correlated materials. At zero exciton density the bilayer Heisenberg system exhibits already interesting magnetism, in the form of the antiferromagnet for small rung coupling turning via an $O(3)$-QNLS quantum phase transition into an "incompressible quantum spin liquid" for larger rung couplings that can be viewed as a continuation of pair singlets ("valence bonds") stacked on the rungs \cite{Chubukov1995}. The natural competitor of the exciton superfluid at finite density is the exciton crystal and one anticipates that due to the strong lattice potential this will tend to lock in at commensurate densities forming exciton "Mott insulators". We will wire this in by taking also the exciton-exciton dipolar interaction into account that surely promotes such orderings. In principle there is the interesting possibility that all these orders may coexist microscopically forming an "antiferromagnetic supersolid" \cite{Zaanen1999}. In this bosonic setting we can address it in a quite controlled manner, but we find that at least for the strongly coupled "small" excitons assumed here this does not happen. The reason is interesting. We already alluded to the dynamical "frustration" associated with the exciton delocalizing in the anti-ferromagnetic spin background, which is qualitatively of the same kind as for the standard "electron" t-J model. At finite densities this turns into a tendency to just phase separate on a macroscopic scale, involving antiferromagnets, exciton crystalline states and high density diamagnetic exciton superfluids, respectively. 

Even though the exciton dipolar repulsion is long-ranged, there is no possibility of frustrated phase separation as suggested for the electronic order in cuprates\cite{Zaanen:1989p1602,Emery:1993p5321,Low:1994p5313,Tranquada:1995p5405,Zhang:2003p5301} because the $1/r^3$ interaction falls off too quickly. However, if one correctly incorporates the full exciton dipolar interaction, a variety of different exciton ordered phase may arise\cite{Rademaker:2013p5682}. Here we restrict ourselves to nearest neighbor repulsion only, which allows for the formation of a checkerboard ordered exciton crystalline state.

It is disappointing that apparently in this system only conventional ground states occur. However, this is actually to a degree deceptive. The Hamiltonian describing the physics at the lattice scale describes a physics where the exciton- and spin motions are "entangled":  the way in which these subsystems communicate gets beyond the notion of just being strongly coupled, since the motions of the exciton motions and the spin dynamics cannot be separated.  By coarse graining this all the way to the static order parameters (the mean fields) an effective decoupling eventually results as demonstrated by the pure ground states. However, upon going "off-shell" this spin-exciton entanglement becomes directly manifest in the form of unexpected and rather counterintuitive effects on the excitation spectrum. A simple example is the zero exciton density antiferromagnet. From the rather controlled linear spin wave self-consistent Born approximation (LSW-SCBA) treatment of the one exciton problem \cite{Rademaker2012EPL} we already know that the resulting exciton spectrum can be completely different from that in a simple semiconductor. We compute here the linearized excitations around the pure antiferromagnet, recovering the LSW-SCBA result in the "adiabatic limit" where the exciton hopping is small compared to the exchange energy of the spin system, which leads to a strong enhancement of the exciton mass. In the opposite limit of fast excitons, the energy scale is recovered but the "Ising-confinement" ladder spectrum revealed by the LSW-SCBA treatment is absent. The reason is clear: in the language of this paper, the couplings between the exciton- and spin-wave modes become very big and these need to be re-summed in order to arrive at an accurate description of the exciton propagator, while our mean-field treatment corresponds with a complete neglect of these exciton-spin interactions.  
  
The real novelty in this regard is revealed in the high density exciton superfluid phase. The spin system forms here a ground state that is a product state of pair-singlets living on the rungs. Besides the superfluid phase modes one expects in addition also the usual massive spin-triplet excitations associated with the (incompressible) singlet vacuum. The surprise is that these are characterized by a dispersion which is in part determined by the {\em superfluid density of the exciton condensate}, as we already announced elsewhere\cite{RademakerMarch2013} for which we present here the details. Counterintuitively, by measuring the spin fluctuations one can in principle determine whether the excitons are condensed in a superfluid.

Let us complete this introduction by specifying the point of departure: the Hamiltonian describing strongly bound excitons propagating through a bilayer Heisenberg spin 1/2 
system. This  model is derived and discussed at length in our earlier papers \cite{Rademaker2012EPL,Rademaker2012arXiv} and here we just summarize the outcome. Due to the strong electron-electron interactions the electronic degrees of freedom are, at electronic half-filling, reduced to spin operators $\mathbf{s}_{i l}$ governed by the bilayer Heisenberg model\cite{Manousakis1991,Chubukov1995}
\begin{equation}
  H_J=
  	J \sum_{\langle ij \rangle,l}\mathbf{s}_{il}\cdot\mathbf{s}_{jl}
  	+J_{\perp}\sum_{i}\mathbf{s}_{i1}\cdot\mathbf{s}_{i2}. \label{HJ}
\end{equation}
The subscript denotes spin operators on site $i$ in layer $l=1,2$. The Heisenberg $H_J$ is antiferromagnetic with $J>0$ and $J_{\perp}>0$. The interlayer exciton can hop around, thereby interchanging places with the spin background. In the strong-coupling limit of exciton binding energies the exciton hopping process is described by the Hamiltonian
\begin{equation}
	H_t = - t \sum_{\langle ij \rangle} | E_j \rangle \left(  | 0 \; 0 \rangle_i \langle 0 \; 0 |_j
			+ \sum_m |1 \; m \rangle_i \langle 1  \; m |_j
			 \right) \langle E_i |.
		\label{ExcitonHop}
\end{equation}
where $| E \rangle$ is the exciton state on an interlayer rung, and $| s \; m \rangle$ represent the rung spin states. Whenever an exciton hops, it effectively exchanges the spin configuration on its neighboring site. This exciton $t-J$ model was derived earlier in Refs. \cite{Rademaker2012EPL, Rademaker2012arXiv}, where the optical absorption was computed in the limit of vanishing exciton density $\langle |E \rangle \langle E | \rangle \rightarrow 0$. In order to study the system with a finite density of excitons, we need to enrich the current $t-J$ model with two extra terms: a chemical potential and an exciton-exciton interaction.

The chemical potential is straightforwardly
\begin{equation}
	H_{\mu} = -\mu \sum_{i} |E_i \rangle \langle E_i |.
	\label{Hmu}
\end{equation}
The exciton-exciton interaction requires more thought. The bare interaction between two interlayer excitons results from their electric dipole moment. Since all interlayer exciton dipole moments are pointing in the same direction the full exciton-exciton interaction is described by a repulsive $1/r^3$ interaction. Hence the interaction strength decays sufficiently fast to avoid the Coulomb catastrophe responsible for frustrated phase separation\cite{Emery:1993p5321,Low:1994p5313}. We consider it reasonable to only include the nearest-neighbor repulsion,
\begin{equation}
	H_{V} = V \sum_{\langle i j \rangle}
		\left( |E_i \rangle \langle E_i | \right)
		\left( |E_j \rangle \langle E_j | \right).
	\label{HV}
\end{equation}
Here $V$ is the energy scale associated with nearest neighbor exciton repulsion. This number can get quite high: given a typical interlayer distance\cite{Imada:1998p2790} of $8 \AA$ and an intersite distance of $4 \AA$ the bare dipole interaction energy is 14 eV. In reality, we expect this energy to be lower due to quantum corrections and screening effects. However, the exciton-exciton interaction scale remains on the order of electronvolts and thus larger than the estimated Heisenberg $J$ and hopping $t$.

Let us finally consider the effects of interlayer hopping of electrons, which leads to the annihilation of excitons,
\begin{equation}
	H_{t_\perp} = - t_\perp \sum_{i} |E_i\rangle \langle 0 \; 0 |_i + h.c.
	\label{InterlayerHopping}
\end{equation}
This term explicitly breaks the $U(1)$ symmetry associated with the conservation of excitons. While this term is almost certainly present in any realistic system, it is a matter of numbers whether it is relevant. In the present case of cuprates, where each layer can be doped by means of chemical substitution, we expect the chemical potential $\mu$ to be significantly larger than the interlayer tunneling $t_\perp$. Consequently, the interlayer hopping is barely relevant. Throughout this publication we will discuss the effects that the inclusion of a small $t_\perp$ will have.

The full model Hamiltonian describing a finite density of excitons in a strongly correlated bilayer is thus
\begin{equation}
	H = H_J + H_t + H_{\mu} + H_V.
	\label{FullH}
\end{equation}

Let us now summarize the layout of our paper. Most of the physics of hard-core excitons on a lattice can be captured using an effective $XXZ$ model, which is studied in section \ref{SecXXZ}. The ground state phase diagram of the full exciton $t-J$ model is derived in section \ref{SecMFT}, using both numerical simulations and analytical mean field theory. The excitations and the corresponding susceptibilities are discussed in section \ref{SecEEx}. We conclude this paper with a discussion on possible further lines of theoretical and experimental research in section \ref{SecConc}.

\section{An effective $XXZ$ model}
\label{SecXXZ}

The Hamiltonian, equation (\ref{FullH}), has five model parameters: $J$, $J_\perp$, $t$, $V$ and $\mu$. However, most properties of the excitons can be understood by considering the problem of hard-core bosons on a lattice. In this section we will argue that the exciton degrees of freedom can be described by an effective $XXZ$ model. Based on some reflections on the mathematical symmetries of the full exciton $t-J$ model, we will describe the properties of this effective $XXZ$ model in subsection \ref{SubSecXXZ1}. We will conclude this section with an outline of the method used to obtain the excitation spectrum of the model.

\subsection{Dynamical and symmetry algebra}
\label{AlgebraSec}
Before characterizing different phases of the model we need to assess the algebraic structure of the exciton $t-J$ model. The set of all operators that act on the local Hilbert space form the \emph{dynamical algebra}, whereas the symmetries of the system are grouped together in the \emph{symmetry algebra}.

To derive the dynamical algebra, it is instructive to start with the bilayer Heisenberg model which has, on each interlayer rung, a $SO(4) \cong SU(2) \times SU(2)$ dynamical algebra\cite{Duin:1997p2301}. Upon inclusion of the exciton hopping term we need more operators, since now the local Hilbert space on an interlayer rung is five-dimensional (four spin states and the exciton). Consider the spin-to-exciton operator $E^+_{sm} \equiv |E \rangle \langle s \; m |$ and its conjugate $E^-_{sm} = (E^+_{sm})^\dagger$. Their commutator reads
\begin{equation}
	[ E^+_{sm} , E^-_{sm} ] = |E \rangle \langle E | - | s\; m \rangle \langle s \; m |
		\equiv 2 E^z_{sm}
\end{equation}
where we have introduced the operator $E^z_{sm}$ to complete a $SU(2)$ algebraic structure. We could set up such a construction for each of the four spin states $| s \; m \rangle$. Under these definitions the exciton hopping term, equation (\ref{ExcitonHop}), can be rewritten in terms of an $XY$-model for each spin state,
\begin{eqnarray}
	H_{t}
	 &=&  -t \sum_{<ij>, sm} 
	 	\left( E^+_{sm,i} E^-_{sm,j} + E^-_{sm,i} E^+_{sm,j} \right) \\
	 &=& -2t \sum_{<ij>, sm} 
	 	\left( E^x_{sm,i} E^x_{sm,j} + E^y_{sm,i} E^y_{sm,j} \right)
\end{eqnarray}
where the sum over $sm$ runs over the singlet and the three triplets. Note that the exciton chemical potential, equation (\ref{Hmu}), acts as an externally applied magnetic field to this $XY$-model, and that the exciton-exciton repulsion, equation (\ref{HV}), can be rewritten as an antiferromagnetic Ising term in the $E^z_{sm}$ operators. The dynamical algebra therefore contains four $SU(2)$ algebras in addition to the $SO(4)$ from the bilayer Heisenberg part. The closure of such an algebra is necessarily $SU(5)$, which is the largest algebra possible acting on the five-dimensional Hilbert space. Hence we need a full $SU(5)$ dynamical algebra to describe the exciton $t-J$ model at finite density. The operators that compose this algebra are enumerated in Appendix \ref{AppendixA}.

From the $XY$-representation of the hopping term one can already deduce that we have four distinct $U(1)$ symmetries associated with spin-exciton exchange. The bilayer Heisenberg model contains two separate $SU(2)$ symmetries, associated with in-phase and out-phase interlayer magnetic order. Therefore the full symmetry algebra of the model is $[SU(2)]^2 \times [U(1)]^4$.

Breaking of the $SU(2)$ symmetry amounts to magnetic ordering, which is most likely antiferromagnetic (and therefore also amounts to a breaking of the lattice symmetry). Each of the $U(1)$ algebras can be broken leading to exciton condensation. Note that next to possible broken continuous symmetries, there also might exist phases with broken translation symmetry. The checkerboard phase, already anticipated in the introduction, is an example of a phase where the lattice symmetry is broken into two sublattices.

\subsection{What to expect: an effective $XXZ$ model}
\label{SubSecXXZ1}
When discussing the dynamical algebra of the exciton $t-J$ model we found that the exciton hopping terms are similar to an $XY$-model. The main reason is that the excitons are, in fact, hard-core bosons and thus allow for a mapping onto pseudospin degrees of freedom. Viewed as such, the exciton-exciton interaction equation (\ref{HV}) is similar to an antiferromagnetic Ising term and the exciton chemical potential equation (\ref{Hmu}) amounts to an external magnetic field in the $z$-direction. Together they form an $XXZ$-model in the presence of an external field,  which has been investigated in quite some detail elsewhere\cite{Neel1936,FisherNelson1974,Landau1981,VanOtterlo1995,Kohno1997,Yunoki2002} as well as in the context of exciton dynamics in cold atom gases\cite{Kantian2007}. 

In order to understand the basic competition between the checkerboard phase and the superfluid phase of the excitons, it is worthwhile to neglect the magnetic degrees of freedom and study first this effective $XXZ$-model for the excitons only. The transition between the checkerboard and superfluid phases is known as the `spin flop'-transition\cite{Neel1936}. Keeping the identification of the exciton degrees of freedom as $XXZ$ pseudospin degrees of freedom in mind, let us review the basics of the $XXZ$ Hamiltonian
\begin{equation}
	H = -t \sum_{\langle i j \rangle} \left( E^x_i E^x_j + E^y_i E^y_j \right)
		- \mu \sum_{i} E^z_i
		+ V \sum_{\langle i j \rangle} E^z_i E^z_j
	\label{XXZmodel}
\end{equation}
where $E^+ = |1 \rangle \langle 0| = E^x + i E^y$ creates a hard-core bosonic particle $| 1 \rangle$ out of the vacuum $|0\rangle$. This model has a built-in competition between $t>0$, which favors a superfluid state, and $V>0$, which favors a crystalline state where all particles are on one sublattice and the other sublattice is empty. The external field or chemical potential $\mu$ tunes the total particle density. The ground state can now be found using mean field theory. It is known that for pseudospin $S=\frac{1}{2}$ models in $(2+1)$D the quantum fluctuations are not strong enough to defeat classical order and therefore we can rely on mean field theory, as supported by exact diagonalization studies\cite{Kohno1997}.

To find the ground state we introduce a variational wavefunction describing a condensate of excitons,
\begin{equation}
	| \Psi \rangle = \prod_i 
		\left( \cos \theta_i e^{i\psi_i} |1 \rangle_i + \sin \theta_i |0 \rangle_i \right).
\end{equation}
The mean-field approximation amounts to choosing $\psi_i$ constant and $\theta_i$ only differing between the two sublattices. We find the following mean-field energy
\begin{eqnarray}
	E/N &=& - \frac{1}{8}tz \sin 2 \theta_A \sin 2\theta_B
	 + \frac{1}{8} Vz \cos 2 \theta_A \cos 2\theta_B
	\nonumber \\ &&
	 - \frac{1}{4} \mu \left( \cos 2 \theta_A + \cos 2\theta_B \right).
\end{eqnarray}
Let's rewrite this in terms of $\overline{\theta} = \theta_A + \theta_B$ and $\Delta \theta = \theta_A - \theta_B$,
\begin{eqnarray}
	E/N &=&  \frac{z}{8} \left(
		(V-t) \cos^2 \Delta \theta + (V+t) \cos^2 \overline{\theta} \right)
	\nonumber \\ &&
	 - \frac{1}{2} \mu \cos \Delta \theta \cos \overline{\theta} - \frac{Vz}{8}.
\end{eqnarray}

\begin{figure}
	\includegraphics[width=\columnwidth]{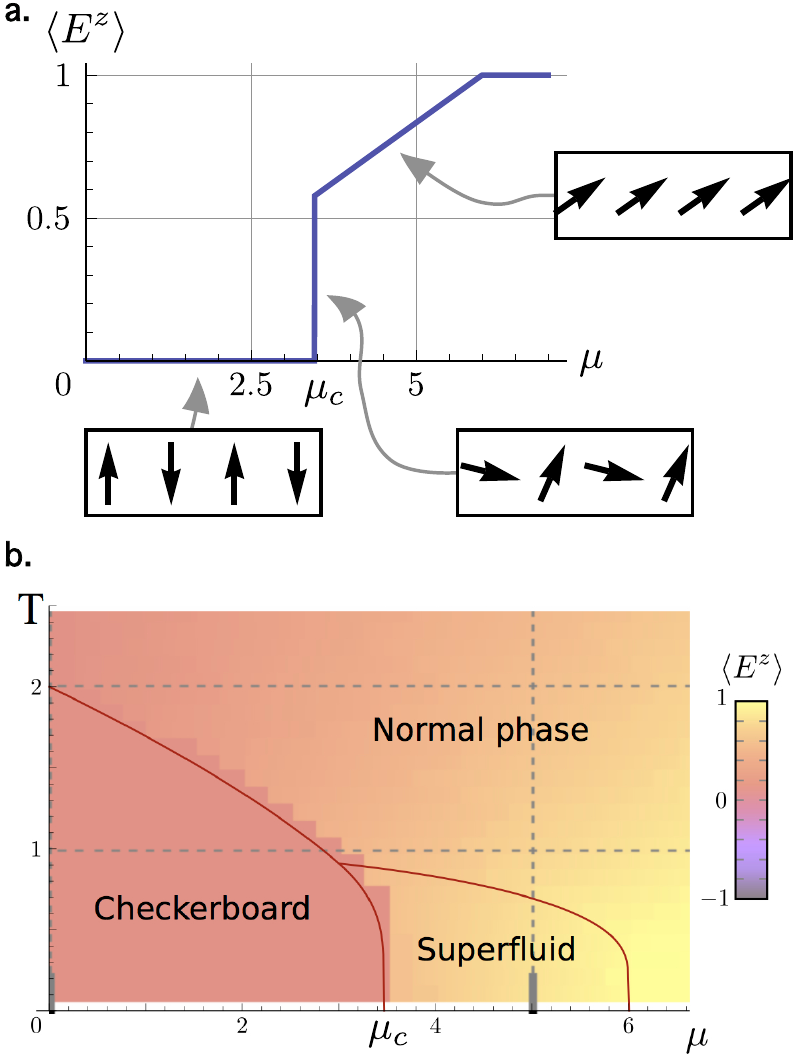}
	\caption{\label{FigXXZ}\textbf{a.} The ground state phase diagram of the $XXZ$ model, equation (\ref{XXZmodel}). The graph shows the mean field particle density $\langle E^z \rangle$ as a function of $\mu$, with model parameters $t=1$ and $V=2t$. One clearly distinguishes the fully polarized phases for large $\mu$, the superfluid phase with a linear $\langle E^z \rangle$ vs $\mu$ dependence and the crystalline checkerboard phase with $\langle E^z \rangle=0$. In between the checkerboard and the superfluid phase a non-trivial first order transition exists, with a variety of coexistence ground states with the same ground state energy. The insets show how the $(E^x,E^z)$-vectors look like in the different phases.
	\textbf{b.} Finite temperature phase diagram of the $XXZ$ model with the same parameters. The background coloring corresponds to a semiclassical Monte Carlo computation of $\langle E^z \rangle$, the solid lines are analytical mean field results for the phase boundaries. We indeed see the checkerboard phase and the superfluid phase, as well as a high-temperature non-ordered `normal' phase.}
\end{figure}

When $|\mu| \geq \frac{1}{2} (V z + zt)$ the ground state is fully polarized in the $z$-direction. This means either zero particle density for negative $\mu$, or a $\rho=1$ for the positive $\mu$ case. Starting from the empty side, increasing $\mu$ introduces a smooth distribution of particles. This phase amounts to the superfluid phase of the excitons. The particle density on the two sublattices is equal and the total density is given by
\begin{equation}
	\rho = \cos^2 \theta = 
	\frac{1}{2} \left( \cos \overline{\theta} + 1 \right) = 
	\frac{1}{2} \left( \frac{2 \mu}{ Vz+zt} + 1 \right).
\end{equation}
At the critical value of the chemical potential
\begin{equation}
	(\mu_c)^2 = \left( \frac{1}{2}z \right)^2 (V - t) (V + t).
\end{equation}
a first order transition occurs towards the checkerboard phase: the spin flop transition. In the resulting phase, which goes under various names such as the antiferromagnetic\footnote{If we associate the presence of a particle with spin up, and the absence with spin down, then the solid phase is identified with an Ising antiferromagnet. However, one should not confuse this with the \emph{actual} antiferromagnetism present in the spin sector of the full exciton $t-J$ model. To avoid confusion, from now on we will use the term 'antiferromagnetism' only when referring to the spin degrees of freedom in the full exciton $t-J$ model.}, solid, checkerboard or Wigner crystalline phase, the sublattice symmetry is broken. The resulting ground state phase diagram is shown in figure \ref{FigXXZ}a, where we also show the dependence of the particle density on $\mu$.

At finite temperatures in $(2+1)$d there can be algebraic long-range order. At some critical temperature a Kosterlitz-Thouless phase transition\cite{KosterlitzThouless} will destroy this long-range order. The topology of the phase diagram however can be obtained using the finite temperature mean field theory for which we need to minimize the mean field thermodynamic potential\cite{Yeomans}
\begin{eqnarray}
	\Phi/N & = & - kT \log \left( 2 \cosh \left( \frac{\beta m}{2} \right) \right)
	+ \frac{1}{2} m \tanh \left( \frac{\beta m}{2} \right) 
	\nonumber \\ &&	
	+ \frac{z}{8} \tanh^2 \left( \frac{\beta m}{2} \right) 
	\nonumber \\ && \; \times
		\left[ (V-t) \cos^2 \Delta \theta + (V+t) \cos^2 \overline{\theta} - V \right]
	\nonumber \\ &&	
	- \frac{\mu}{2} \tanh \left( \frac{\beta m}{2} \right) 
		\cos \Delta \theta \cos \overline{\theta}.
\end{eqnarray}
Expectation values are 
\begin{equation}
	\langle S^x_{i \in A} \rangle = \frac{1}{2} \sin 2 \theta_A  \tanh \left( \frac{\beta m}{2} \right),
\end{equation}
and the parameter $m$ needs to be determined self-consistently. The resulting phase diagram is shown in figure \ref{FigXXZ}b, which is of the form discussed by Fisher and Nelson\cite{FisherNelson1974}.

The first order quantum phase transition at $\mu_c$ turns out to be non-trivial, a point which is usually overlooked in the literature. A trivial first order transition occurs when there are two distinct phases with exactly the same energy. In the case presented here, there is a infinite set of mean field order parameters all yielding different phases yet still having the same energy. A simple analytic calculation shows that the energy of the ground state at the critical point is $E_c = -Vz/8$. Now rewrite the mean field parameters $\rho_A$ and $\rho_B$ into a sum and difference parameter
\begin{eqnarray}
	\rho &=& \frac{1}{2} ( \rho_A + \rho_B), \\
	\Delta_\rho & = & \frac{1}{2} (\rho_A - \rho_B ).
\end{eqnarray}
For each value of $\Delta_\rho$ with $|\Delta_\rho| \leq (1/2)$ we can find a value of $\rho$ such that the mean field energy is exactly $-Vz/8$. 

This has interesting consequences. If one can control the density instead of the chemical potential around a first order transition, in general phase separation would occur between the two competing phases. From the mean field considerations above it is unclear what would happen in a system described by the $XXZ$ Hamiltonian, equation (\ref{XXZmodel}). All phases would be equally stable, at least on the mean field level, and every phase may occur in regions of any size. Such a highly degenerate state may be very sensible to small perturbations. We consider it an interesting open problem to study the dynamics of such a highly degenerate system, and whether this degeneracy may survive the inclusion of quantum corrections.

In the introduction we mentioned the existence of interlayer hopping, equation (\ref{InterlayerHopping}). Qualitatively the $t_\perp$ is irrelevant, which can be seen in the $XXZ$ pseudospin language where it takes the form of a tilt of the magnetic field in the $x$-direction,
\begin{equation}
	H_{t_\perp}  = - t_\perp \sum_i E^x_i.
\end{equation} 
As a result the phase diagram is shifted but not qualitatively changed. The effect of the $t_\perp$ on the excitation spectrum is briefly discussed in section \ref{SFphasesection}.

\subsection{Excitations of the $XXZ$ model}
\label{XXZexc}

Of direct experimental relevance are the elementary excitations of a phase. The dispersion of these excitations can be computed using the `equations of motion'-method based on the work of Zubarev\cite{Zubarev1960}. We present the formalities of this method in Appendix \ref{AppendixC}. In this subsection we briefly show the essence of this technique, applied to the $XXZ$ model. Later, in section \ref{SecEEx}, we will compute the excitations for the full exciton $t-J$ model.

The key ingredients of this Zubarev-approach are the Heisenberg equations of motion,
\begin{eqnarray}
	i \partial_t E^+_i &=&
		-t \sum_\delta E^z_i E^+_{i+\delta} + \mu E^+_i 
		- V \sum_\delta E^+_i E^z_{i+\delta}, \\
	i \partial_t E^-_i &=&
		t \sum_\delta E^z_i E^-_{i+\delta} - \mu E^-_i 
		+ V \sum_\delta E^+_i E^z_{i+\delta}, \\
	i \partial_t E^z_i &=&
		- \frac{1}{2} t \sum_\delta \left( E^+_i E^-_{i+\delta} - E^-_i E^+_{i+\delta} \right),
\end{eqnarray}
where $\delta$ runs over all nearest neighbors. These equations cannot be solved exactly, and one relies on the approximation controlled by the mean field vacua. That is, we neglect fluctuations of the order parameters, so that products of operators on different sites are replaced by\cite{Zubarev1960,Oles:2000p5088}
\begin{equation}
	A_i B_j \rightarrow \langle A_i \rangle B_j + A_i \langle B_j \rangle
\end{equation}
where $\langle \ldots \rangle$ denotes the mean field expectation value. By such a decoupling the Heisenberg equations of motion become a coupled set of linear equations which can be solved easily. In the homogeneous phase we thus obtain, after Fourier transforming,
\begin{eqnarray}
	\omega_k E^+_k &=& 
		- \frac{1}{2} tz \left( \cos 2 \theta \gamma_k E^+_k + \sin 2 \theta E^z_k \right)  + \mu E^+_k 
	\nonumber \\ &&
		- \frac{1}{2} Vz \left( \cos 2 \theta E^+_k + \sin 2 \theta \gamma_k E^z_k \right) \\
	\omega_k E^-_k &=&
		\frac{1}{2} tz \left( \cos 2 \theta \gamma_k E^-_k + \sin 2 \theta E^z_k \right)  - \mu E^-_k 
	\nonumber \\ &&
		+ \frac{1}{2} Vz \left( \cos 2 \theta E^-_k + \sin 2 \theta \gamma_k E^z_k \right) \\
	\omega_k E^z_k &=&
		- \frac{1}{4} tz \sin 2 \theta (1 - \gamma_k) \left( E^+_k - E^-_k \right).
\end{eqnarray}
We find an analytical expression for the excitations in the superfluid phase,
\begin{eqnarray}
	\omega_k & =& 
		\frac{1}{2} zt \sqrt{1 - \gamma_k} 
		\sqrt{1 - \gamma_k (1-2\rho)^2 + \frac{4V}{t} \gamma_k (1-\rho) \rho } 
	\nonumber \\
		&=& \frac{1}{2} z t \sqrt{ \rho (1-\rho) ( 1 + V/t)} \; |k| + \ldots
		\label{XXZsuperfluidSpeed}
\end{eqnarray}
where $\gamma_k = \frac{1}{2} (\cos k_x + \cos k_y )$. For small momenta this excitation has a linear dispersion, conform to the Goldstone theorem requiring a massless excitation as a result of the spontaneously broken $U(1)$ symmetry. Exactly at $\mu=\mu_c$ the dispersion reduces to $\omega_k = zt \sqrt{ 1 - \gamma_k^2}$, hence the gap at $k=(\pi,\pi)$ closes thus signaling a transition towards the checkerboard phase.

At the critical point and in the checkerboard phase, we need to take into account the fact that expectation values of operators differ on the two sublattices. The Heisenberg equations of motion now reduce to six (instead of three) linear equations, which can be straightforwardly solved. For now we postpone the discussion on the dispersion of elementary excitations to section \ref{SecEEx}, where the full exciton $t-J$ model will be considered using the technique discussed here.

\section{Ground state phase diagram}
\label{SecMFT}
In the previous section we have seen that the effective $XXZ$ model predicts the existence of both an exciton superfluid phase and a checkerboard phase, separated by a first order transition. Now we derive the ground state phase diagram for the full exciton $t-J$ model given by equation (\ref{FullH}).

We will proceed along the same lines as in the previous section, starting with a variational wavefunction. Numerical simulation of this wavefunction creates an unbiased view on the possible inhomogeneous and homogeneous ground state phases. This serves as a basis to further analyze the phase diagram with analytical methods. The analytical mean field theory also allows us to characterize the three homogeneous phases: the antiferromagnet, the superfluid and the checkerboard crystal. Finally, combining the numerical and analytical mean field results we obtain the ground state phase diagram, see figure \ref{FinalPhaseDiagramFig}.

\subsection{Variational wavefunction for the exciton $t-J$ model}
\label{SSVariational}
Recall that the local Hilbert space consists of four spin states $| s \; m \rangle$ and the exciton state $| E \rangle$. We therefore propose a variational wavefunction consisting of a product state of a superposition of all five states on each rung. For the spin states we take the $SO(4)$ coherent state\cite{Duin:1997p2301}
\begin{eqnarray}
	| \Omega_i \rangle &=&
		- \frac{1}{\sqrt{2}} \sin \chi_i \sin \theta_i e^{- i \phi_i } | 1 \; 1 \rangle_i
	\nonumber \\ &&
		+ \frac{1}{\sqrt{2}} \sin \chi_i \sin \theta_i e^{ i \phi_i } | 1 \; -1 \rangle_i
	\nonumber \\ &&
		+ \sin \chi_i \cos \theta_i | 1 \; 0 \rangle_i
		- \cos \chi_i | 0 \; 0 \rangle_i
\end{eqnarray}
which needs to be superposed with the exciton state,
\begin{equation}
	| \Psi_i \rangle = \sqrt{\rho_i} e^{i \psi_i} | E_i \rangle + \sqrt{1-\rho_i} | \Omega_i \rangle 
\end{equation}
to obtain the total variational (product state) wavefunction
\begin{equation}
	| \Psi \rangle = \prod_i | \Psi_i \rangle.
	\label{VariationalWavefunction}
\end{equation}
This full wavefunction acts as ansatz for the numerical simulations. Note that the homogeneous phases can be described by this wavefunction with the parameters $\chi, \theta, \phi, \psi$ and $\rho$ only depending on the sublattice. Given this wavefunction, the expectation value of a product of operators on different sites decouples, $\langle A_i B_j \rangle = \langle A_i \rangle \langle B_j \rangle$. The only nonzero expectation values of spin operators are for $\widetilde{\mathbf{S}}_i = \mathbf{s}_{i1} - \mathbf{s}_{i2}$ and it equals
\begin{equation}
	\langle \Omega_i | \widetilde{\mathbf{S}}_i | \Omega_i \rangle
		= 	\sin 2\chi_i
			\begin{pmatrix}
				\sin \theta_i \cos \phi_i \\
				\sin \theta_i \sin \phi_i \\
				\cos \theta_i
			\end{pmatrix}
		= \sin 2 \chi_i \; \hat{\mathbf{n}}_i
\end{equation}
where $\hat{\mathbf{n}}_i$ is the unit vector described by the angles $\theta$ and $\phi$. This variational wavefunction therefore assumes interlayer N\'{e}el order of magnitude $\sin 2 \chi_i$, which enables us to correctly interpolate between the perfect N\'{e}el order at $\chi = \pi/4$ and the singlet phase $\chi=0$ present in the bilayer Heisenberg model. The exciton density at a rung $i$ is trivially given by $\rho_i$.

\subsection{Simulated annealing}
Given the variational wavefunction, we can use simulated annealing to develop an unbiased view on the possible mean field ground state phases. Therefore we start out with a lattice with on each lattice site the variables $\theta_i$, $\chi_i$, $\phi_i$, $\psi_i$ and $\rho_i$ and with periodic boundary conditions. The energy of a configuration is
\begin{eqnarray}
	E  &=&  \frac{1}{2} J \sum_{<ij>} 
			(1-\rho_i) (1-\rho_j) \sin 2 \chi_i \sin 2 \chi_j \; 
			\hat{n}_i \cdot \hat{n}_j
	\nonumber \\ &&
		- J_\perp \sum_i (1 - \rho_i) \cos^2 \chi_i
		- \mu \sum_i \rho_i + V \sum_{<ij>} \rho_i \rho_j
	\nonumber \\ &&
		- \frac{1}{2} t \sum_{<ij>} \sqrt{\rho_i (1-\rho_i)\rho_j (1-\rho_j)} \cos ( \psi_i  - \psi_j)
	\nonumber \\ &&
			\; \times \left(
			\cos \chi_i \cos \chi_j + \sin \chi_i \sin \chi_j
			\; \hat{n}_i \cdot \hat{n}_j \right)
		\label{SCEnergy}
\end{eqnarray}
We performed standard Metropolis Monte Carlo updates of the lattice with fixed total exciton density. The fixed total exciton density is imposed as follows: if during an update the exciton density $\rho_i$ is changed, the exciton density on one of the neighboring sites is corrected such that the total exciton density remains constant.

\begin{figure}
	\includegraphics[width=\columnwidth]{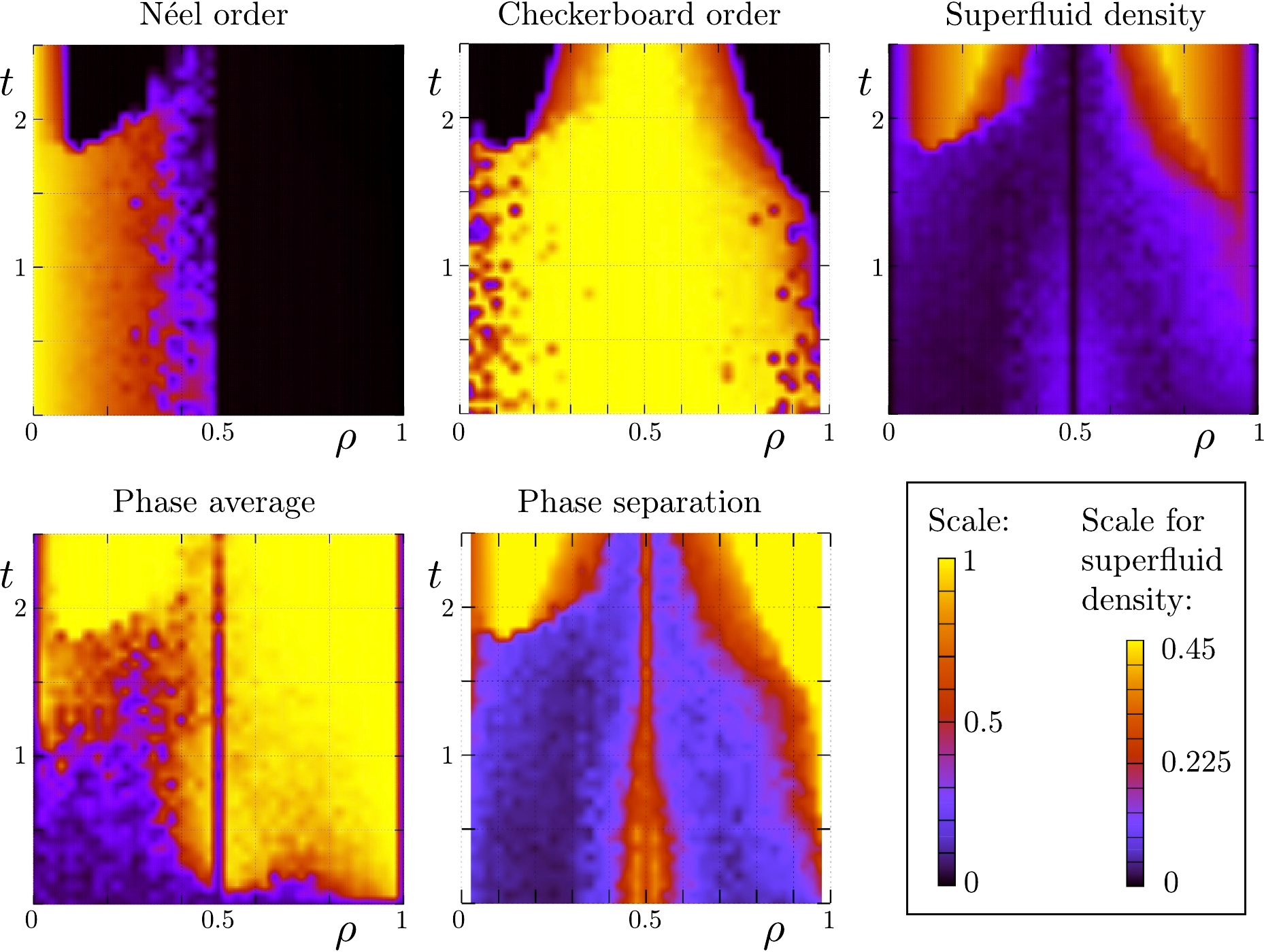}
	\caption{\label{MCresults}Results from the semi-classical Monte Carlo simulations. Here shown are color plots, with on the horizontal axes the exciton density $\rho$ and on the vertical axes the hopping parameter $t$ (in eV). Other parameters are fixed at $J = 125$ meV, $\alpha = 0.04$ and $V = 2$ eV. The five measurements shown here are the N\'{e}el order parameter, equation (\ref{FirstMCformula}), the checkerboard order parameter, equation (\ref{CBMCformula}), the superfluid density, equation (\ref{SFdensityformula}), the phase coherence, equation (\ref{PhaseMCF}), and the ratio signaling phase separation according to equation (\ref{LastMCformula}), 0 means complete phase separation, 1 means no phase separation. Notice that the prominent line at $\rho=0.5$ signals the checkerboard phase.}
\end{figure}

The main results of the simulation are shown in figure \ref{MCresults}, for various values of the hopping parameter $t$ and exciton density $\rho$. We performed the computations on a $10 \times 10$ lattice. Notice that even though true long-range order does not exist in two dimensions, the correlation length of possible ordered phases is larger than the size of our simulated lattice. The other parameters are fixed at $J = 125$ meV, $\alpha = 0.04$ and $V = 2$ eV. The Heisenberg couplings $J = 125$ meV and $\alpha = 0.04$ are obtained from measurements of undoped YBCO-samples\cite{Imada:1998p2790,Tranquada:1989p5209}, which we consider to be qualitatively indicative of all strongly correlated electron bilayers. The dipolar coupling is estimated at 2 eV, following our discussion in the introduction. 

For each value of $\rho$ and $t$ we started at a high temperature $T=0.1$ eV, to slowly reduce the temperature to $10^{-5}$ eV while performing a full update of the whole lattice 10 million times. We expect that by such a slow annealing process we obtain the true ground state of equation (\ref{SCEnergy}), devoid of topological defects. Once we arrive at the low temperature state, we performed measurements employing 200.000 full updates of the system.

We measured six different order parameter averages:
\begin{itemize}
	\item The N\'{e}el order parameter defined by
	\begin{equation}
		\mathrm{Neel} = \left| \left| \frac{1}{N} \sum_i (-1)^i (1 - \rho_i) \sin 2 \chi_i \hat{\mathbf{n}}_i \right| \right|
		\label{FirstMCformula}
	\end{equation}
	where we first sum over all spin vectors and then take the norm.
	\item The checkerboard order, defined as the difference in exciton density between the sublattices divided by the maximal difference possible. The maximal difference possible equals $\mathrm{Min}(\overline{\rho}, 1-\overline{\rho})$, so
	\begin{equation}
		\mathrm{Checkerboard} = \frac{ \frac{1}{N} \sum_i (-1)^i \rho_i }{\mathrm{Min}(\overline{\rho}, 1-\overline{\rho})}.
		\label{CBMCformula}
	\end{equation}
	\item The superfluid density is given by the expectation value of the exciton operator. Here we do not make a distinction between singlet exciton condensation or triplet exciton condensation. Therefore 
	\begin{equation}
		\mathrm{Superfluid \; density} = \frac{1}{N} \sum_i \sqrt{ \rho_i ( 1- \rho_i)}.
		\label{SFdensityformula}
	\end{equation}
	\item Now the superfluid density is not the only measure of the condensate, we can also probe the rigidity of the phase $\psi$. Therefore we sum up all the phase factors on all sites,
	\begin{equation}
		\mathrm{Phase \; average} = \left| \frac{1}{N} \sum_i e^{i \psi_i} \right|.
		\label{PhaseMCF}
	\end{equation}
	If the phase is disordered, this sum tends to zero. On the other hand, complete phase coherence in the condensate phase implies that this quantity equals unity.
	\item Finally, we considered a measure of phase separation between the checkerboard phase and the superfluid phase. If the exciton condensate and the checkerboard phase are truly coexisting, then the maximal superfluid density attainable would be 
	\begin{eqnarray}
		\mathrm{Max\; SF \; density} =
		\frac{1}{2} \sqrt{ (\overline{\rho} + \Delta_\rho) (1 - \overline{\rho} - \Delta_\rho)} 
		\nonumber \\
		- \frac{1}{2} \sqrt{ (\overline{\rho} - \Delta_\rho) (1 - \overline{\rho} + \Delta_\rho)}
	\end{eqnarray}
	where $\Delta_\rho = \frac{1}{N} \sum_i (-1)^i \rho_i$. If there is phase separation however, the actual superfluid density is less than this maximal density. Therefore we also measured the ratio
	\begin{equation}
		\mathrm{Ratio} = \frac{\mathrm{Superfluid \; density}}{\mathrm{Max\; SF \; density}}
		\label{LastMCformula}
	\end{equation}
	to quantify the extent of phase separation. When this ratio is less than $1$ this indicates phase separation.
\end{itemize}

\begin{figure}
	\includegraphics[width=\columnwidth]{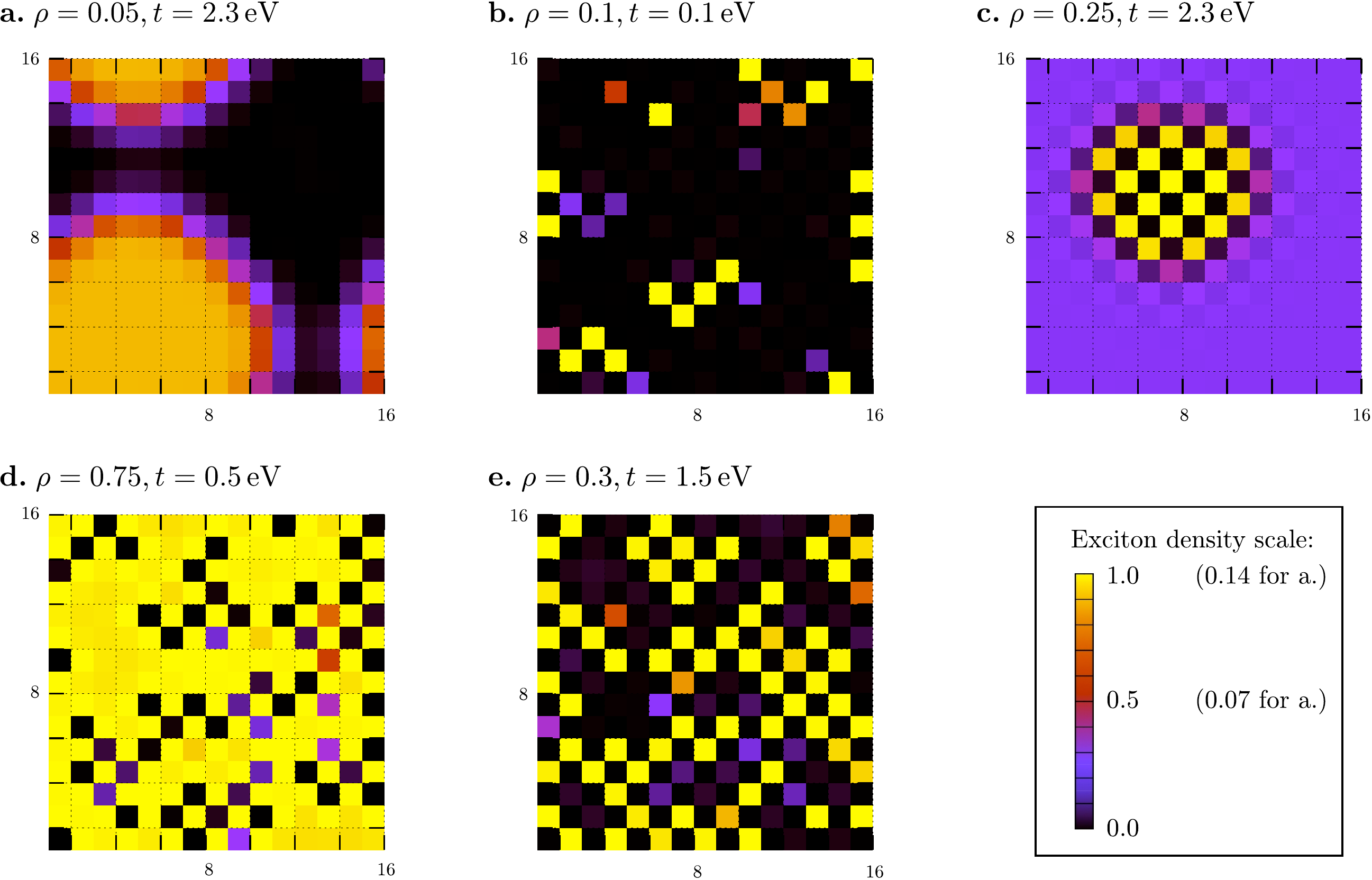}
	\caption{\label{MCdensityplots} Typical configurations for the exciton density per site, obtained in the Monte Carlo simulation on a $16 \times 16$ square lattice. The color scale indicates the exciton density. All five figures have model parameters $J = 125$ meV, $\alpha = 0.04$ and $V = 2$ eV.  \textbf{a:} Separation between the antiferromagnetic phase (without excitons, hence shown black) and the exciton condensate with smooth exciton density ($\rho = 0.05$, $t = 2.3$ eV). \textbf{b:} Separation between checkerboard-like localized excitons and an antiferromagnetic background ($\rho = 0.1$, $t = 0.1$ eV). \textbf{c:} Separation between the checkerboard phase and a low density exciton condensate ($\rho = 0.25$, $t = 2.3$ eV). \textbf{d:} Separation between the checkerboard phase and a high density exciton condensate ($\rho = 0.75$, $t = 0.5$ eV). \textbf{e:} The region where antiferromagnetic order, checkerboard order and the exciton condensate are all present ($\rho=0.3$, $t=1.5$).}
\end{figure}

\begin{figure}
	\includegraphics[width=\columnwidth]{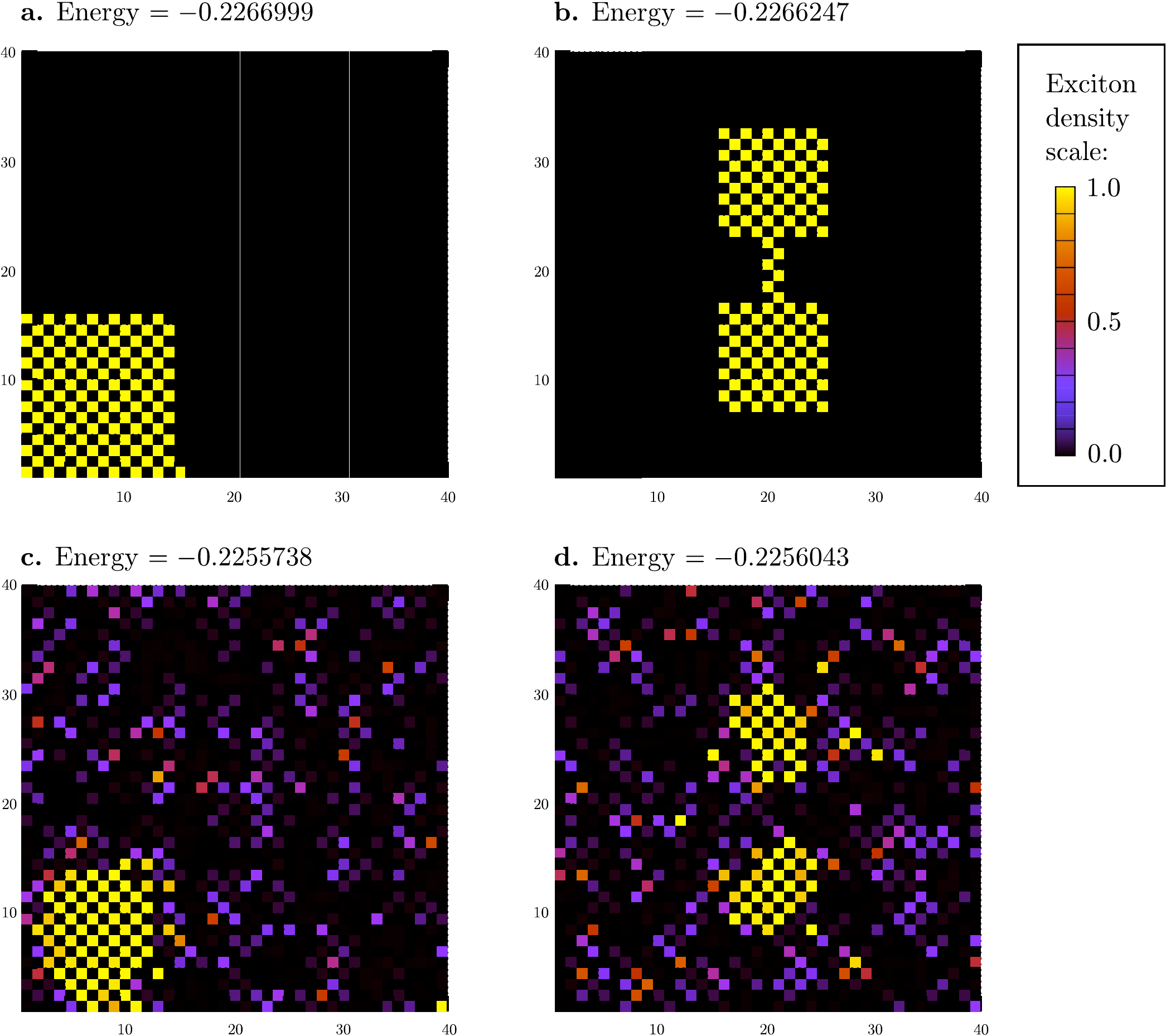}
	\caption{\label{PhaseSepFig} Different exciton configurations with their respective energies on a $40 \times 40$ lattice, to show whether there is macroscopic phase separation. The model parameters are $t=0.5$ eV, $J = 125$ meV, $\alpha = 0.04$, $V = 2$ eV and $\rho = 0.06625$. Yellow indicates the presence of excitons, and in the black regions there is antiferromagnetic order. \textbf{a:} The lowest energy state is the one with complete macroscopic phase separation. \textbf{b:} More complicated phase separation, such as the halter form depicted here, are higher in energy. \textbf{c:} Starting at high temperatures with the configuration a, we slowly lowered the temperature. The resulting configuration shown here is a local minimum. \textbf{d:} Using the same slow annealing as for c starting from configuration b. The local energy minimum obtained this way is lower in energy than the configuration c. We conclude that even though macroscopic phase separation has the lowest energy, there are many local energy minima without macroscopic phase separation.}
\end{figure}

The results for a full scan for the range $0 < \rho < 1$ and $0 < t < 2.5$ eV are shown in figure \ref{MCresults}. In figures \ref{MCdensityplots} and \ref{PhaseSepFig} we have displayed typical exciton density configurations for various points in the phase diagram. In combination these results suggest that there are three homogeneous phases present in the system: the antiferromagnet at low exciton densities, the exciton superfluid at high exciton hopping energies and the checkerboard crystal at half-filling of excitons. However, for most parts of the phase diagram the competition between the three phases appears to result in phase separation.

Let us investigate the phase separation in somewhat more detail. In our earlier work we found that the motion of an exciton in an antiferromagnetic background leads to dynamical frustration\cite{Rademaker2012EPL,Rademaker2012arXiv}. In other words: excitons do not want to coexist with antiferromagnetism. The introduction of a finite density of excitons will therefore induce phase separation. For large $t$, we find macroscopic phase separation between the antiferromagnet and the exciton superfluid, see figure \ref{MCdensityplots}a. At low exciton kinetic energy the excitons will crystallize in a checkerboard pattern as can be seen in figure \ref{MCdensityplots}b. 

Close to half-filling the role of the dipole repulsion $V$ becomes increasingly relevant. The first order `spin flop' transition we discussed in section \ref{SubSecXXZ1} implies that there will be phase separation between the superfluid and the checkerboard order. Figures \ref{MCdensityplots}c and d show this phase separation. Finally there is a regime where the condensate, the checkerboard order and the N\'{e}el order are all present. However, given the dynamical frustration on the one hand and the spin-flop transition on the other hand, we again predict phase separation. A typical exciton configuration in this parameter regime is shown in figure \ref{MCdensityplots}e.

These simulated annealing results suggest that phase separation dominates the physics of this exciton system. To check whether the numerics are reliable we inspected directly the energies of the various homogeneous mean field solutions, using the Maxwell construction for phase separated states. The constructed phase separated configurations and their energies are shown in figure \ref{PhaseSepFig}. The lowest energy configuration (\ref{PhaseSepFig}a) has macroscopic phase separation between the checkerboard and the antiferromagnetic phase. Intermediate states with one blob of excitons (\ref{PhaseSepFig}c) are slightly higher in energy than states with two blobs of excitons (\ref{PhaseSepFig}d). However, even though macroscopic phase separation has the lowest energy, configurations with more blobs have more entropy. Consequently for any nonzero temperatures complete macroscopic phase separation is not the most favorable solution. This is indeed seen in the numerical simulations: annealing leads to high-entropy states such as figure \ref{PhaseSepFig}d rather than to the lowest energy configuration.

We thus conclude that the dominant phases are the antiferromagnet, the superfluid and the checkerboard. The competition between these three phases leads to phase separation in most parts of the phase diagram. The unbiased Monte Carlo simulations show the direction in which further analytical research should be directed: we will use mean field theory to characterize the three homogeneous phases.

\subsection{Mean field theory and characterization of the phases}
\label{SSGrandcanonical}

Given the fact that we are dealing with a hard-core boson problem, we know that mean field theory is qualitatively correct. A remaining issue is whether one can tune the exciton chemical potential rather than the exciton density in realistic experiments. Since we are prescient about the many first-order phase transitions in this system, we will perform the analysis with a fixed exciton density (the canonical ensemble). Using the Maxwell construction and the explicit $\mu$ vs. $\rho$ relations, we can transform back to the grand-canonical ensemble.

\begin{figure}
	\includegraphics[width=\columnwidth]{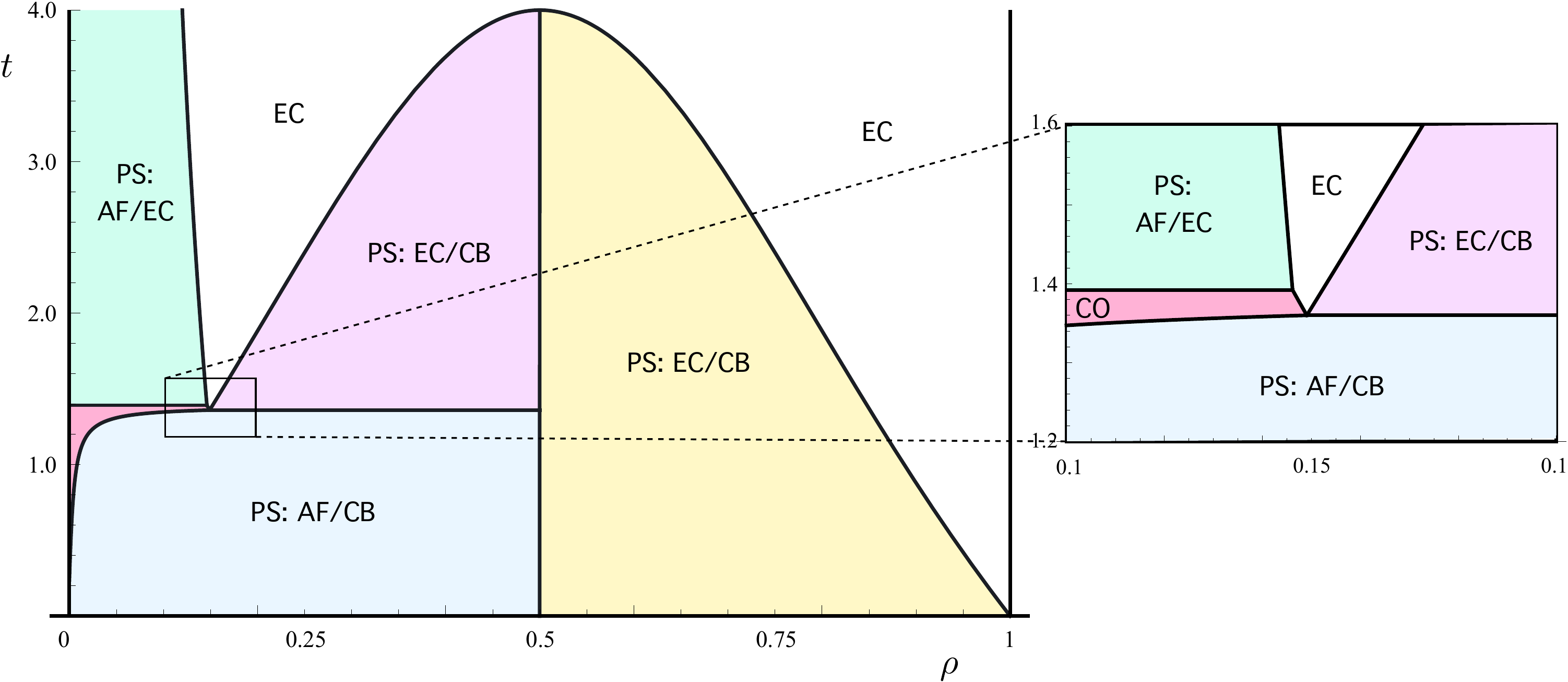}
	\caption{\label{CanonicalDiagram} The canonical mean-field phase diagram for typical values of $J = 125$ meV, $\alpha = 0.04$ and $V = 2$ eV whilst varying $t$ and the exciton density $\rho$. In the absence of exciton, at $\rho=0$, we have the pure antiferromagnetic N\'{e}el phase (AF). Exactly at half-filling of excitons ($\rho=1/2$) and small hoping energy $t<2V$ we find the checkerboard phase (CB) where one sublattice is filled with excitons and the other sublattice is filled with singlets. For large values of $t$ we find the singlet exciton condensate (EC), given by the wavefunction $\prod_i \left( \sqrt{\rho} \hat{E}^+_{00,i} + \sqrt{1 - \rho} \right) | 0\; 0 \rangle_i$. The coexistence of antiferromagnetism and superfluidity for small $\rho$ and $t$ is an artifact of the mean field theory. Conform the Monte Carlo results of figure \ref{MCresults}, for most parts of the phase diagram phase separation (PS) is found.}
\end{figure}

The numerical simulations suggest that the only solutions breaking translational symmetry invoke two sublattices,
\begin{equation}
	\rho_i = \left\{ \begin{array}{ll} \rho_A & \; i \in A \\ \rho_B & \; i \in B \end{array}
		\right.
\end{equation}
and so forth for $\chi$, $\theta$, $\psi$ and $\phi$. This broken translational symmetry allows for the antiferromagnetic and exciton checkerboard order. Evaluation of the energy $E = \langle \Psi | H | \Psi \rangle$ of the variational wavefunction, equation (\ref{VariationalWavefunction}), directly suggests that we can set $\theta = \psi = \phi= 0$ on all sites.\footnote{By setting $\theta = \phi =0$ we restrict the spin vectors to be pointing in the $\pm z$ direction only. Since we anticipate magnetic ordering we have the freedom to choose the direction of the ordering. Similar arguments hold for the choice $\psi=0$; when breaking the $U(1)$ symmetry associated with exciton condensation we are free to choose the phase direction.} We are left with four parameters $\rho_A, \rho_B, \chi_A$ and $\chi_B$, and as it turns out it will be more instructive to rewrite these in terms of sum and difference variables,
\begin{eqnarray}
	\overline{\rho} & = & \frac{1}{2} (\rho_A + \rho_B), \\
	\Delta_\rho & = & \frac{1}{2} (\rho_A - \rho_B), \\
	\overline{\chi} & = & \chi_A + \chi_B, \\
	\Delta_\chi & = & \chi_A - \chi_B .
\end{eqnarray}
The mean field energy per site is now given by
\begin{eqnarray}
	 E/N &=& \frac{1}{8} J z \left( (1 - \overline{\rho})^2 - \Delta_\rho^2 \right)
			( \cos 2 \Delta_\chi - \cos 2 \overline{\chi} )
	\nonumber \\ &&
		- \frac{1}{2} J_\perp \left[ (1-\overline{\rho})(\cos \overline{\chi} \cos \Delta_\chi + 1) 
	\right. \nonumber \\ && \left. \;
			+ \Delta_\rho \sin \overline{\chi} \sin \Delta_\chi \right]
	\nonumber \\ &&
		- \frac{1}{4} zt \sqrt{((1 - \overline{\rho})^2 - \Delta_\rho^2)(\overline{\rho}^2 - \Delta_\rho^2)}
			\cos \Delta_\chi
	\nonumber \\ &&
		- \mu \overline{\rho} + \frac{1}{2} z V (\overline{\rho}^2 - \Delta_\rho^2)
	\label{MeanFieldE}
\end{eqnarray}
which has to be minimized for a fixed average exciton density $\overline{\rho}$ with the constraint $|\Delta_\rho| \leq \min(\overline{\rho}, 1-\overline{\rho})$. The resulting mean field phase diagram for typical values of $J, J_\perp$ and $V$, and for various $t, \overline{\rho}$, is shown in figure \ref{CanonicalDiagram}.

\subsubsection{Antiferromagnetic phase}

As long as the exciton density is set to zero, the mean field ground state is given by the ground state of the bilayer Heisenberg model,
\begin{equation}
	\rho = 0, \; \overline{\chi} = 0 
		\; \mathrm{and} \;
		\cos \Delta_\chi = \frac{J_\perp}{Jz} \equiv \alpha.
\end{equation}	
The N\'{e}el order is given by 
\begin{equation}
	\frac{1}{N} \sum_i (-1)^i \langle \widetilde{S}^z_i \rangle = \sqrt{1 - \alpha^2}
\end{equation}
and the energy of the antiferromagnetic state is
\begin{equation}
	E = -\frac{1}{4} J z ( 1 + \alpha)^2.
\end{equation}
The introduction of excitons in an antiferromagnetic background leads to dynamical frustration effects which disfavors the coexistence of excitons and antiferromagnetic order\cite{Rademaker2012EPL,Rademaker2012arXiv}. In fact, the numerical simulations already ruled out coexistence of superfluidity and antiferromagnetism.

\subsubsection{Exciton condensate} For large exciton hopping energy $t$ it becomes more favorable to mix delocalized excitons into the ground state. Due to the bosonic nature of the problem this automatically leads to exciton condensation. The delocalized excitons completely destroy the antiferromagnetic order and the exciton condensate is described by a superposition of excitons and a singlet background,
\begin{equation}
	| \Psi \rangle = 
		\prod_i
		\left( \sqrt{\rho} | E_i \rangle + \sqrt{1-\rho} | 0 \; 0 \rangle_i \right).
\end{equation}

Here we wish to emphasize the ubiquitous coupling to light of the superfluid. The dipole matrix element allows only spin zero transitions, and since the exciton itself is $S=0$ the dipole matrix element is directly related to the superfluid density,
\begin{eqnarray}
	\langle \sum_{\sigma} c^\dagger_{i1 \sigma} c_{i2\sigma} \rangle
	&=& \langle E | \left( c^\dagger_{1 \uparrow} c_{2\uparrow}
		+ c^\dagger_{1 \downarrow} c_{2\downarrow} \right) | 0 \; 0 \rangle
	\nonumber \\
	&=& \frac{1}{\sqrt{2}} \sqrt{\rho (1-\rho)} 
		\langle \uparrow \downarrow_1 \; 0_2 | 
	\nonumber \\ && 
		\left( c^\dagger_{1 \uparrow} c_{2\uparrow}
		+ c^\dagger_{1 \downarrow} c_{2\downarrow} \right)
		\left( 
			| \uparrow_1 \; \downarrow_2 \rangle - | \downarrow_1 \; \uparrow_2 \rangle \right)
	\nonumber \\
	&=& \sqrt{2 \rho (1-\rho)}
\end{eqnarray}
The dipole matrix element thus acts as the order parameter associated with the superfluid phase. In most bilayer exciton condensates, such as the one in the quantum Hall regime\cite{Eisenstein:2004p4770}, this order parameter is also nonzero in the normal phase because of interlayer tunneling of electrons. One can therefore not speak strictly about spontaneous breaking of $U(1)$ symmetry in such systems; there is already explicit symmetry breaking due to the interlayer tunneling. In strongly correlated electron systems the finite $t_\perp$ is small compared to the chemical potential $\mu$. As discussed in the introduction, the Mott insulating bilayers now effectively allow for spontaneous $U(1)$ symmetry breaking, and the above dipole matrix element acts as a true order parameter. Note that the irrelevance of interlayer hopping $t_\perp$ implies that this order parameter is, unfortunately, not reflected in photon emission or interlayer tunneling measurements.

The exciton condensate is a standard two-dimensional Bose condensate. The $U(1)$ symmetry present in the $XY$-type exciton hopping terms is spontaneously broken and we expect a linearly dispersing Goldstone mode in the excitation spectrum, reflecting the rigidity of the condensate. We will get back to the full excitation spectrum in section \ref{SecEEx}.

The energy of the singlet exciton condensate is
\begin{equation}
	E = -J_\perp - \frac{( \mu + \frac{1}{4} zt - J_\perp)^2}{zt + 2Vz}
\end{equation}
and the exciton density is given by
\begin{equation}
	\overline{\rho} = 2\frac{\mu + \frac{1}{4} zt - J_\perp}{zt + 2Vz}.
\end{equation}

\subsubsection{Checkerboard phase}
Whenever the exciton hopping is small, the introduction of excitons into the system leads to the `spin flop' transition towards the checkerboard crystalline phase. As shown in the context of the $XXZ$ model, this phase implies that one sublattice is completely filled with excitons and the other sublattice is completely empty. On the empty sublattice, any nonzero $J_\perp$ will guarantee that the singlet spin state has the lowest energy. Hence the average exciton density is here $\overline{\rho}=\Delta_\rho =1/2$ and the energy of the checkerboard phase is given by
\begin{equation}
	E = - \frac{1}{2} J_\perp - \frac{1}{2} \mu.
\end{equation}
It is interesting to note that the checkerboard phase is in fact similar to a Bose Mott insulator: with the new doubled unit cell we have one exciton per unit cell. The nearest neighbor dipole repulsion now acts as the `on-site' energy preventing extra excitons per unit cell.

\subsubsection{Coexistence of antiferromagnetism and exciton condensate}

Within the analytical mean field theory set by equation (\ref{MeanFieldE}) there exists a small region where antiferromagnetism and the exciton condensate coexist. There the energy of the homogenous coexistence phase is lower than the energy of macroscopic phase separation of the antiferromagnet and the condensate, as obtained using the Maxwell construction. However, within numerical simulations we found no evidence of coexistence. Instead, we found microscopic phase separation, which hints at a possible complex inhomogeneous phase. We therefore conclude that the homogeneous mean field theory discussed here is insufficient to find the true ground state.

\subsubsection{Exciton Mott insulator}
Finally, when the exciton density is unity we have a system composed of excitons only. In the parlance of hard-core bosons this amounts to a exciton Mott insulator. This rather featureless phase is adiabatically connected to a standard electronic band insulator: the system is now composed of two layers where each layer has an even number of electrons per unit cell. The energy of the exciton Mott insulator is, trivially
\begin{equation}
	E = -\mu + \frac{1}{2} Vz.
\end{equation}

\begin{widetext}
\subsection{Phase separation}
\label{SSCanonical}

In this mean field theory most of the phase transitions are first order, with the exciton density varying discontinuously along the transition. The critical values of $\mu$ or $t/J$ for the first order transitions are
\begin{eqnarray}
	\mu_{c, \mathrm{AF}\rightarrow \mathrm{CB}} & =& 
		\frac{1}{2} J z (1 + \alpha^2) 
		\label{CriticalMuAFCB} \\
	\mu_{c, \mathrm{CB}\rightarrow \mathrm{EI}} & =& 
		Vz + J_\perp \\
	(t/J)_{c, \mathrm{AF}\rightarrow \mathrm{EC}} & =& 
		2 (1+\alpha^2) - 4 \frac{\mu}{Jz} 
		+ 2 \sqrt{(1-\alpha^2)
		\left(4\frac{\mu}{Jz}- (1+\alpha)^2 - 2\frac{V}{J} \right)}
		\label{CriticalMuAFEC} \\
	(t/J)_{c, \mathrm{CB}\rightarrow \mathrm{EC}} & = &
		4 \sqrt{ \left( \frac{\mu}{Jz} - \alpha \right)\left( \frac{V}{J} + \alpha - \frac{\mu}{Jz}\right)} \\
	(t/J)_{c, \mathrm{CO}\rightarrow \mathrm{CB}} & = &
		\frac{2 \alpha^2}{2\frac{\mu}{Jz} - 1} - 2 \alpha 
		+ \sqrt{
			\left( 1 - \frac{\alpha^2}{2\frac{\mu}{Jz} - 1} \right)
			\left( 2 \left(\frac{V}{J} + \alpha - \frac{\mu}{Jz} \right)
				- \frac{\alpha^2}{2\frac{\mu}{Jz} - 1} \right)}.
\end{eqnarray}
The transitions towards the coexistence region from the antiferromagnet or the condensate are second order. Additionally, the transition from the condensate to the exciton Mott insulator is second order. The critical values of $t/J$ or $\mu$ at these second order transitions are
\begin{eqnarray}
	(t/J)_{c, \mathrm{AF}\rightarrow \mathrm{CO}} & = &
		\frac{2Jz (1+\alpha) - 4 \mu}{J_\perp} \\
	(t/J)_{c, \mathrm{EC}\rightarrow \mathrm{CO}} & = &
		1 - \frac{2\mu}{Jz} + \sqrt{ (1+8\alpha) + \left(\frac{2\mu}{Jz} \right)^2
		- 4 \left( 3 \frac{\mu}{Jz} - \frac{2V}{J} (1-\alpha)\right) }\\
 \mu_{c, \mathrm{EC}\rightarrow \mathrm{EI}} & =& 
		J_\perp + \frac{1}{4} zt + Vz.
\end{eqnarray}
\end{widetext}
The subscripts indicate the phases: antiferromagnetic phase (AF), coexistence phase (CO), exciton condensate (EC), exciton Mott insulator (EI), checkerboard phase (CB).

For any nonzero $\alpha$ the first order transitions from the antiferromagnetic or coexistence phase towards the checkerboard phase are `standard' in the sense that at the critical value of $\mu$ there are only two mean field states with equal energy. This is also true for the transitions from the antiferromagnet to the exciton condensate except at a single point. At the tricritical point
\begin{eqnarray}
	t_c &=&
		2J \sqrt{2V/J-1} \\
	\mu_c &=&
		J_\perp - \frac{1}{4} zt + \frac{1}{2} J z (1-\alpha) \sqrt{ 2V/J + t/J}
\end{eqnarray}
separating the coexistence phase, the antiferromagnetic phase and the exciton condensate, we can set the parameters $\overline{\chi}=0$, $\Delta_\rho=0$ and $\Delta_\chi$ given by the value in the coexistence phase. Now the energy becomes independent of the exciton density $\overline{\rho}$. Similarly, at the critical value of
\begin{equation}
	\mu_c = J_\perp + \frac{1}{2} Vz \pm \frac{1}{4} \sqrt{ (2Vz)^2 - (zt)^2}
	\label{CriticalMuECCB}
\end{equation}
describing the transition between the checkerboard phase to the singlet exciton condensate, we can choose the mean field parameters $\overline{\chi}=0$, $\Delta_\chi = 0$ and
\begin{equation}
	\Delta_\rho = \frac{1}{\sqrt{2}}
		\sqrt{(1-2\rho + 2 \rho^2) - \frac{2V |1-2\rho|}{\sqrt{4V^2-t^2}} }.
\end{equation}
With these parameters, the energy becomes independent of $\rho$.

This implies that the mean field theory predicts highly degenerate states at the critical values of $\mu$, similar to the one we found in the $XXZ$ model. The phase separation that thus occurs can be between an infinite set of possible ground states that have all a different exciton density. Coincidentally, the numerical simulations indicate that around the two `degenerate' critical points indeed all the three phases are present. While the macroscopic phase separated state might have the lowest energy, figure \ref{PhaseSepFig} suggests that more complicated patterns of phase separation are likely to occur. The degeneracy of the critical points on the level of mean fields theory might be responsible for richer physics in these special regions of the phase diagram.

\subsection{Conclusion}

\begin{figure}
	\includegraphics[width=\columnwidth]{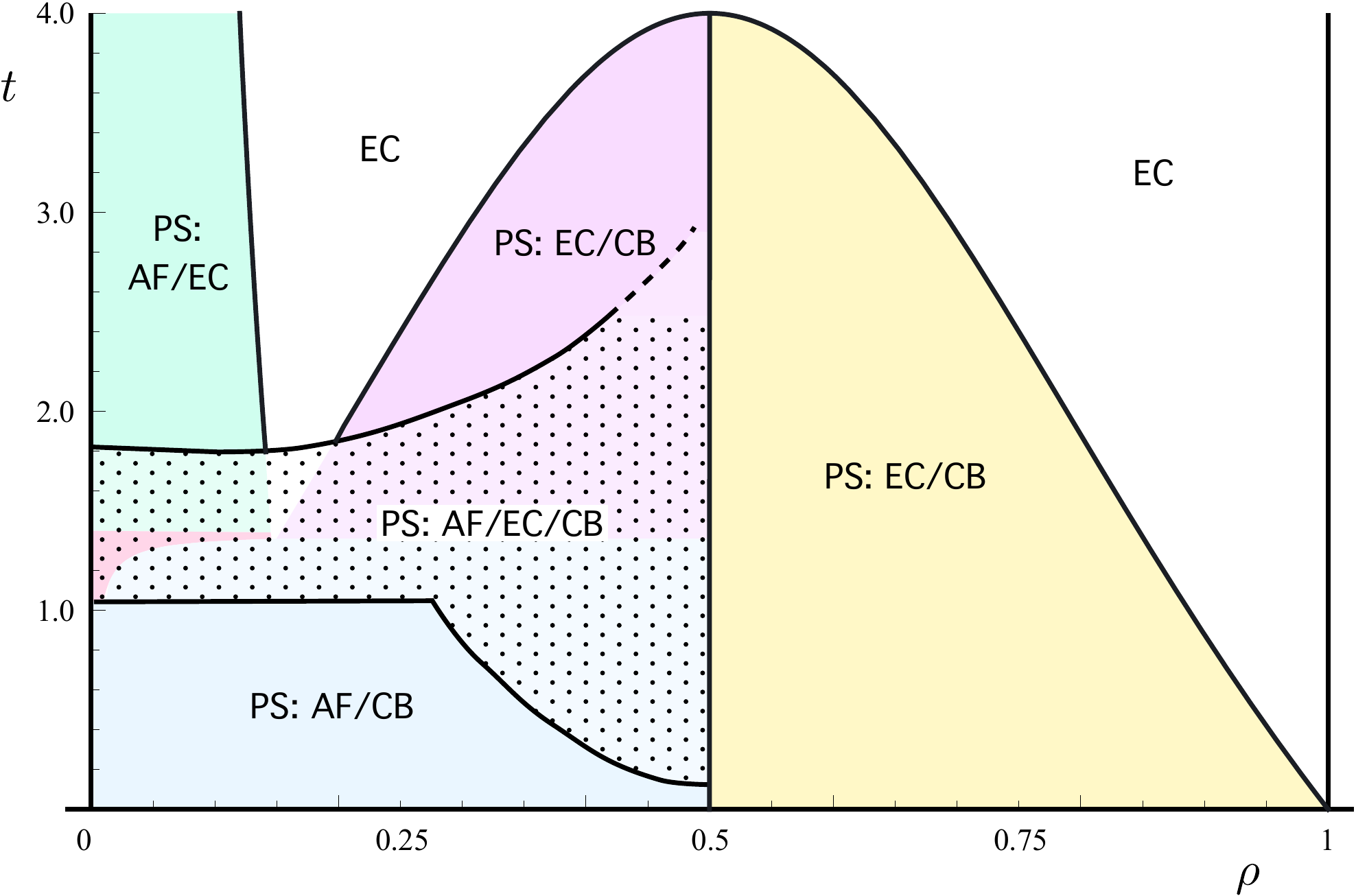}
	\caption{\label{FinalPhaseDiagramFig}The canonical ground state phase diagram of the exciton $t-J$ model, which is a combination of the semi-classical Monte Carlo result and the mean field computations. In the background we have put the mean field phase diagram of figure \ref{CanonicalDiagram}, whilst the lines show the phase diagram as obtained from the Monte Carlo simulations. The dotted area represents phase separation between the condensate, antiferromagnetic and checkerboard order. Furthermore: EC means exciton condensate, CB means checkerboard phase, AF means antiferromagnetism and PS stands for phase separation.}
\end{figure}

\begin{figure}
	\includegraphics[width=\columnwidth]{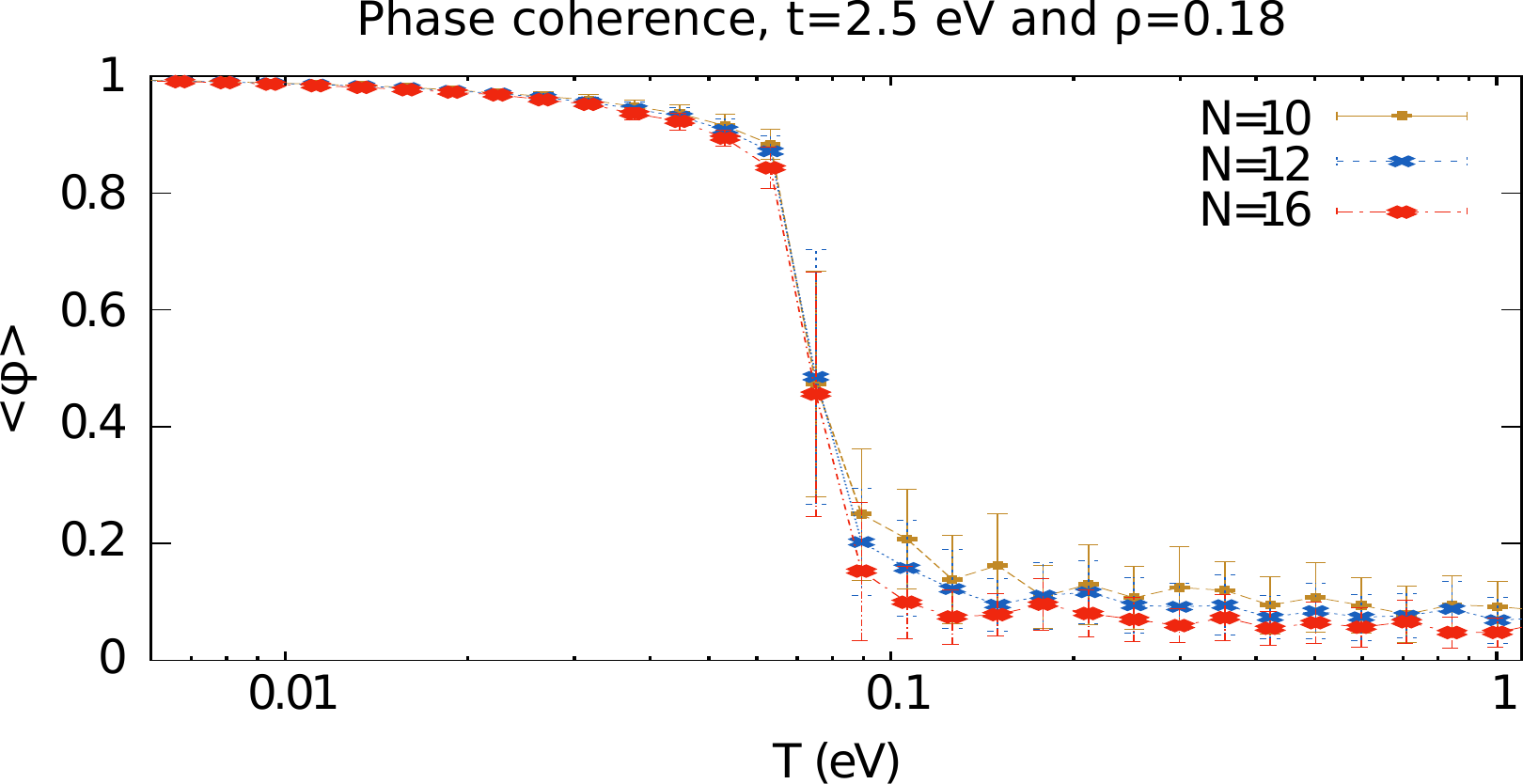}
	\caption{\label{FiniteTresults} Finite temperature graph of the phase coherence in the exciton condensate region of the phase diagram. Here $t=2.5$ eV and $\rho=0.18$ and the other parameters are the same as in a. A clear transition is observed at around 0.06 eV, which amounts to a transition temperature of about 700 Kelvin. }
\end{figure}

Combining the simulated annealing results of figure \ref{MCresults} with the analytical mean field results of figure \ref{CanonicalDiagram} we arrive at the definitive mean field phase diagram of the exciton $t-J$ model in figure \ref{FinalPhaseDiagramFig}. There are three main phases: the antiferromagnet at zero exciton density, the checkerboard crystal at exciton density $\rho=1/2$ and the superfluid at high hopping energy $t$. For most parts of the phase diagram, phase separation between these three phases occurs in any possible combination. The competition between these three phases leads generally to macroscopic phase separation.

Finally, within the limitations of the semi-classical Monte Carlo approach we deduce an estimate of the transition temperature towards the superfluid state. Given a typical point in the phase diagram where the exciton condensate exists, at $t=2.5$ eV and $\rho = 0.18$, we find a Kosterlitz-Thouless transition temperature of approximately 700 Kelvin, see figure \ref{MCresults}c. This number should be taken not too seriously, as the exciton $t-J$ model might not be applicable at such high temperatures given possible exciton dissociation. Additionally, at high temperatures the electron-phonon coupling becomes increasingly important, which we neglect in our exciton $t-J$ model. Nonetheless, our estimate suggests that exciton superfluidity may extend to quite high finite temperatures.

\section{Collective modes and susceptibilities}
\label{SecEEx}

Each phase of the excitons in the strongly correlated bilayer has distinct collective modes, that are in principle measurable by experiment. In order to obtain the dispersions of the collective modes we employ the technique of the Heisenberg equations of motion, introduced in the context of the $XXZ$ model in section \ref{XXZexc} and further formalized in Appendix \ref{AppendixC}. In the case of the exciton $t-J$ model the set of equations is larger and analytical solutions can in general not be obtained. Whenever this is the case we compute the dispersions numerically.

Quantities of direct experimental relevance are the dynamical susceptibilities. We are for instance interested in the absorptive part of the dynamical magnetic susceptibility, defined by
\begin{equation}
	\chi_{S}'' (q, \omega) = 
		\sum_n \langle \psi_0 | \widetilde{S}^-(-q) | n \rangle 
			\langle n | \widetilde{S}^+ (q) | \psi_0 \rangle
			\delta(E_n - \omega)
	\label{chiT}
\end{equation}
Here $| \psi_0 \rangle$ is the ground state of the system and $|n \rangle$ are the excited states with energy $E_n$. It appears unlikely that bilayer exciton systems can be manufactured in bulk form which is required for neutron scattering, while there is a real potential to grow these using thin layer techniques. Therefore the detection of the dynamical spin susceptibility forms a realistic challenge for resonant inelastic X-ray scattering (RIXS)\cite{Ament:2010p5208} measurements with its claimed sensitivity for interface physics\cite{Dean2012}. 

Furthermore we are interested in the charge dynamical susceptibility
\begin{equation}
	\chi_{E}'' (q, \omega) = 
		\sum_n \langle \psi_0 |E^-_{00} (-q) | n \rangle 
			\langle n | E^+_{00} (q) | \psi_0 \rangle
			\delta(E_n - \omega).
	\label{chiE}
\end{equation}
which is directly related to the polarization propagator. We use the operator $E_{00} (q)$ because this amounts to the interlayer dipole matrix element. Therefore, this charge dynamical susceptibility expresses the excitonic excitations. It can be observed by optical absorption experiments\cite{Basov:2005p1689} at $q=0$. Finite wavelength measurements may be obtained using the aforementioned RIXS\cite{Ament:2010p5208} technique, or using electron energy loss spectroscopy (EELS)\cite{Schnatterly1979,Wang:1996p5177}. The method we use to compute the susceptibilities, based on the Heisenberg equations of motion method, is also described in Appendix \ref{AppendixC}.

The three dominant phases we encountered in our mean field analysis will have distinct magnetic and optical responses. Let us briefly summarize our main findings with respect to the collective excitations. The results for the antiferromagnetic phase are shown in figures \ref{AFSpinWaves} to \ref{AFAntiAdiabatic}. This limit of vanishing exciton density has been studied with in far greater rigor than our current Zubarev method is capable of\cite{Rademaker2012EPL,Rademaker2012arXiv}. We can therefore compare the results of the Zubarev method with a full resummation of spin-exciton interactions using the self-consistent Born approximation (LSW-SCBA). It turns out that for small exciton kinetic hopping $t$ the non-interacting equations-of-motion method yields reliable results. For large $t$ one needs the full SCBA code to correctly reproduce the dynamical frustration effects of excitons in the antiferromagnetic background.

The collective modes of the exciton condensate are shown in figures \ref{EC-exciton} and \ref{EC-spin}. Due to the absence of dynamical frustration and the presence of a spin-gap we expect that these results survive in a fully interacting computation. In fact, here the modes of the simple hard-core boson system discussed in section \ref{SecXXZ} can be used as a template. Just as for the phase diagram, the qualitative features of $XXZ$ model are still of relevance for the more complicated $t-J$ model. Nonetheless, in this condensate phase the interplay between excitonic and magnetic degrees of freedom gives rise to a rather counterintuitive effect. We find that the \emph{exciton superfluid density} can be detected directly in a measurement of the \emph{magnetic excitations}, as we already announced elsewhere\cite{RademakerMarch2013}.

In contrast, in the checkerboard crystalline phase the spin and exciton degrees of freedom are once again decoupled. In the remainder of this section we will elaborate further on these results for each phase separately. Throughout the following discussion, the model parameters are $J = 125$ meV, $\alpha = 0.04$, $V=2$ eV and a varying $t$ and $\rho$. In order to visualize the susceptibilities we have convoluted $\chi''$ with a Lorentzian of width $0.04$ eV. The color scale of the susceptibility plots is in arbitrary units.

\subsection{Antiferromagnetic phase: a single exciton}
\label{SecAFmodes}

\begin{figure}
	\includegraphics[width=\columnwidth]{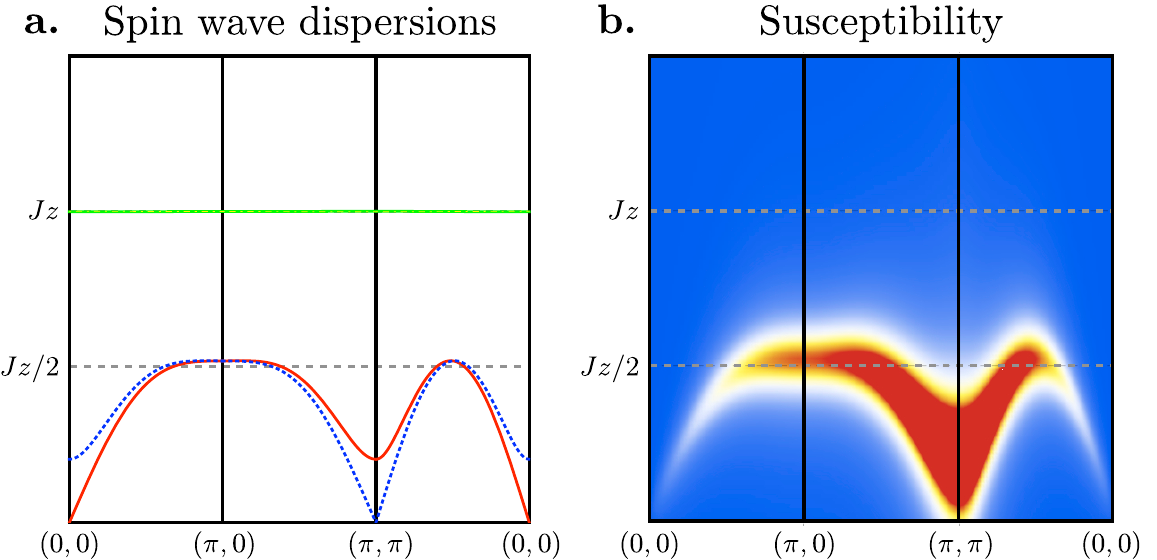}
	\caption{\label{AFSpinWaves} The spin wave dispersions (\textbf{a.}) and the dynamical magnetic susceptibility (\textbf{b.}) in the antiferromagnetic phase. 
	In this phase, the spin wave dispersions are not influenced by exciton dynamics. As is known from previous studies, there are two transversal spin waves and two longitudinal spin waves\cite{Chubukov1995,Rademaker2012arXiv}. The transversal spin waves are gapless around either $\Gamma$ (solid red line) or the $M$ point (dotted blue line). The longitudinal spin waves, which are associated with interlayer fluctuations (solid green line), are nearly flat and have a gap of order $Jz$. The dynamic magnetic susceptibility (\textbf{b.}) only shows one transversal spin wave. These results and all subsequent figures are obtained using $J=125$ meV and $\alpha = 0.04$, as is expected for the undoped bilayer cuprate YBCO \cite{Tranquada:1989p5209}.}
\end{figure}

\begin{figure}
	\includegraphics[width=\columnwidth]{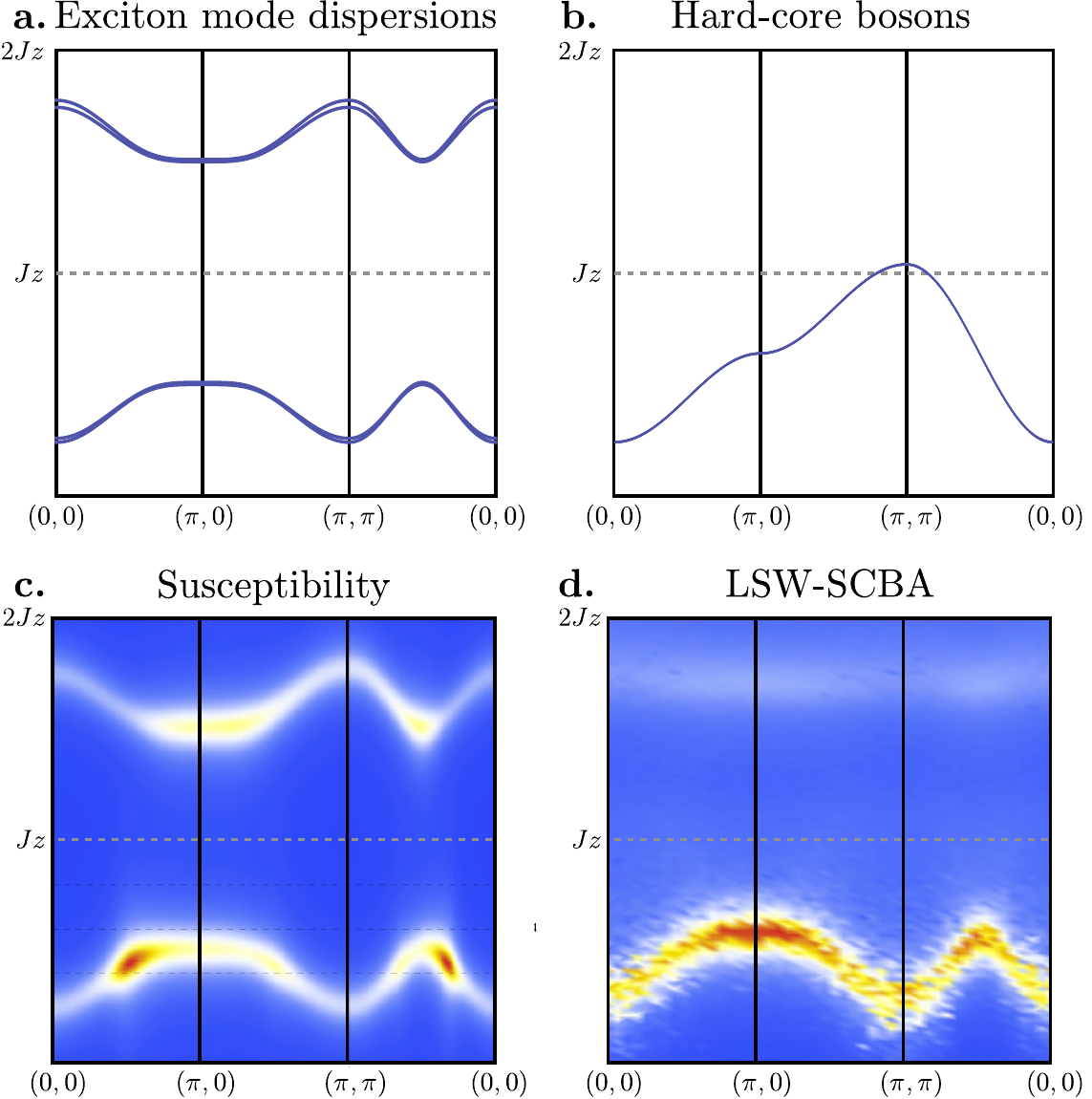}
	\caption{\label{AFAdiabatic}
	The exciton modes in the antiferromagnetic phase in the adiabatic regime $t \ll J$. Here we have chosen $t=0.1$ eV, $J=125$ meV and $\alpha=0.04$.
	Within the equations of motion picture there are four exciton modes (\textbf{a.}), which come in pairs of two with a small interlayer splitting. Due to the antiferromagnetic order the exciton bands are renormalized with respect to a free hard-core boson (\textbf{b.}). The susceptibility corresponding to the free exciton motion (\textbf{c.}) is verified by the fully interacting LSW-SCBA results (\textbf{d.}). This is to be expected: in the adiabatic regime spins react much faster than the exciton motion and the exciton still moves freely dressed by a spin polaron, reducing its bandwidth to order $t^2/J$.}
\end{figure}

\begin{figure}
	\includegraphics[width=\columnwidth]{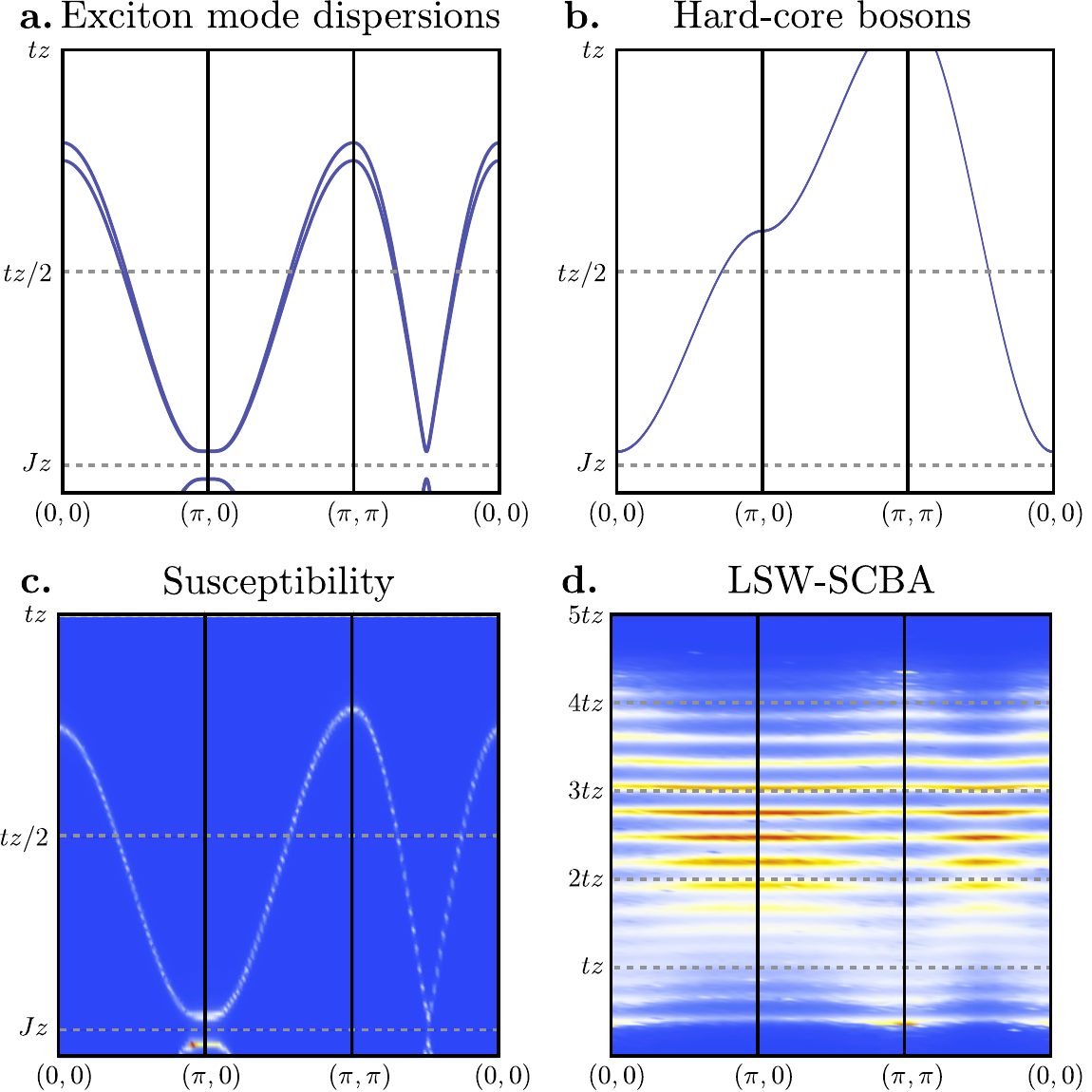}
	\caption{\label{AFAntiAdiabatic}
	The exciton modes in the antiferromagnetic phase in the antiadiabatic regime $t \gg J$. Here we have chosen $t=2$ eV, $J=125$ meV and $\alpha=0.04$.. 
	Just like in figure \ref{AFAdiabatic} we find four exciton bands (\textbf{a.}), renormalized with respect to the free hard-core boson results (\textbf{b.}). However, upon inclusion of the interaction the free susceptibility (\textbf{c.}) gets extremely renormalized (\textbf{d.}). The large exciton kinetic energy together with the relatively spin dynamics create an effective potential for the exciton: the exciton becomes localized and the confinement generates a ladder spectrum. Note that thus in the antiadiabatic regime the free results (\textbf{a.}, \textbf{c.}) cannot be trusted.}
\end{figure}

In the limit of zero exciton density we recover the well-known bilayer Heisenberg physics\cite{Chubukov1995}. As discussed in section \ref{SSGrandcanonical}, the spins tend to order antiferromagnetically. The excitations spectrum thus contains a Goldstone spin wave with linear dispersion around $\Gamma$ and a similar mode centered around $(\pi,\pi)$. In addition, the bilayer nature is reflected in the presence of two longitudinal spin waves with a gap of order $Jz$ and a narrow bandwidth of order $J_\perp$. The excitation spectrum and the corresponding magnetic dynamical susceptibility is shown in figure \ref{AFSpinWaves}. Since the spin modes of the bilayer antiferromagnet are independent of any exciton degrees of freedom, we will not discuss these any further.

The dynamics of an isolated exciton in an antiferromagnetic background has been studied extensively by means of a linear spin-wave self-consistent Born approximation technique (LSW-SCBA)\cite{Rademaker2012EPL,Rademaker2012arXiv}. The non-interacting equations of motion method used in this paper, amounts to the complete neglect of exciton-spin interactions, while these are on the foreground of the (resummed) LSW-SCBA computation. However, the mere existence of LSW-SCBA results allows us to compare it with our current non-interacting calculations. Let us therefore first go through the LSW-SCBA results. There we need to distinguish between two limits: the adiabatic limit with $t \ll J$ shown in figure \ref{AFAdiabatic}, and the anti-adiabatic limit where $t \gg J$ shown in figure \ref{AFAntiAdiabatic}.

Consider a single exciton in an antiferromagnetic background. Now if this exciton hops to a neighboring site, it will leave behind two spins that are ferromagnetically aligned with their neighbors. This process is called dynamical frustration and limits severely the motion of an exciton. In the adiabatic limit ($t \ll J$) this causes the exciton bandwidth to be drastically reduced to an order $t^2/J$. In addition, the magnetic background acts as a confining potential leading to small but detectable ladder states at higher energies.

At the other hand, in the anti-adiabatic regime $t \gg J$ exciton hopping will destroy the antiferromagnetic order as it will be surrounded by a cloud of frustrated spins. The quasiparticle picture completely breaks down and the spectral weight of the exciton is redistributed to a wide incoherent spectral bump. The ladder spectrum arising from the effective confinement will still be visible, though smeared out.

The equations-of-motion method however ignores the effects of spin-exciton interactions such as dynamical frustration. It treats the excitons as well-defined quasiparticles. As such we can already guess beforehand that the non-interacting results will be reliable in the adiabatic regime. Indeed, in the equations-of-motion method we find four exciton modes corresponding to either the singlet $E^+_{00}$ or $m=0$ triplet exciton $E^+_{10}$ operator, just as in the LSW-SCBA. When $\alpha \rightarrow 0$ we can write out an analytical expression for the non-interacting dispersions,
\begin{equation}
	\omega_{k,\pm} = 
		\mu \pm \frac{1}{2} 
			\sqrt{ (Jz)^2  + \left( \frac{1}{2} z t \gamma_k \right)^2 }.
	\label{ExcDispersion}
\end{equation}
where each branch is twofold degenerate. This degeneracy is lifted when $\alpha \neq 0$, leading to a splitting of order $\alpha$ which is largest around $\Gamma$ and $M$.

In the limit of $t \ll J$ the dispersions, equation (\ref{ExcDispersion}), indeed result in an effective exciton bandwidth of order $t^2/J$, conform the fully interacting theory as can be seen in figure \ref{AFAdiabatic}. The natural question then arises: how is it possible that in the present non-interacting theory the exciton bandwidth depends on the spin parameter $J$? For sure, the effective exciton model introduced in section \ref{SecXXZ} has no such renormalization as is shown in figure \ref{AFAdiabatic}. There the exciton bandwidth fully depends on $zt$.

However, it is important to realize that the exciton operators $E^+_{s0,i}$ do not commute with the antiferromagnetic order parameter operator $\widetilde{S}^z_i$. As a result the mean field energy of exciting an exciton is shifted either up or down (depending on the sublattice) yielding a gap between the two exciton branches of $\mathcal{O}(Jz)$. Now for small $t$, propagation of the exciton requires that one has to 'pay' the energy shift $Jz$ to move through both sublattices. As a result the effective hopping is reduced by a factor $t/J$. Therefore the exciton bandwidth renormalization, seen in the full LSW-SCBA, is already present at the mean field level.

For large $t/J$ however we will pay a price for the convenience of the non-interacting equations of motion method. At the mean field level one still expects the dispersions to be described by equation (\ref{ExcDispersion}). However, upon inclusion of the interaction corrections this picture breaks down completely. The bandwidth of the non-interacting exciton is of order $zt$, whereas in the interacting theory an incoherent ladder spectrum of the same width arises. Thus for large $t/J$ the non-interacting results cannot be trusted. However, this only applies to the antiferromagnetic phase due to the presence of dynamical frustration. In general, it appears that the non-interacting results are qualitatively correct in the absence of gapless modes that need to be excited in order for an exciton to move. This condition is naturally met for the other two phases. We therefore expect that exciton-spin interactions only lead to qualitative changes in the antiferromagnetic phase.

By simple selection rules one can already conclude that the singlet exciton mode couples to light. As a consequence this is the mode that is visible in the charge dynamical susceptibility, which is related to the polarization propagator. The exciton excitations are shown in figures \ref{AFAdiabatic}d (for $t<J$) and \ref{AFAntiAdiabatic}d (for $t>J$).

Finally, note that at the transition from the antiferromagnetic phase to the checkerboard phase the gap in the exciton spectrum vanishes at $(\pi,\pi)$.

\subsection{Superfluid phase}
\label{SFphasesection}

\begin{figure}
	\includegraphics[width=\columnwidth]{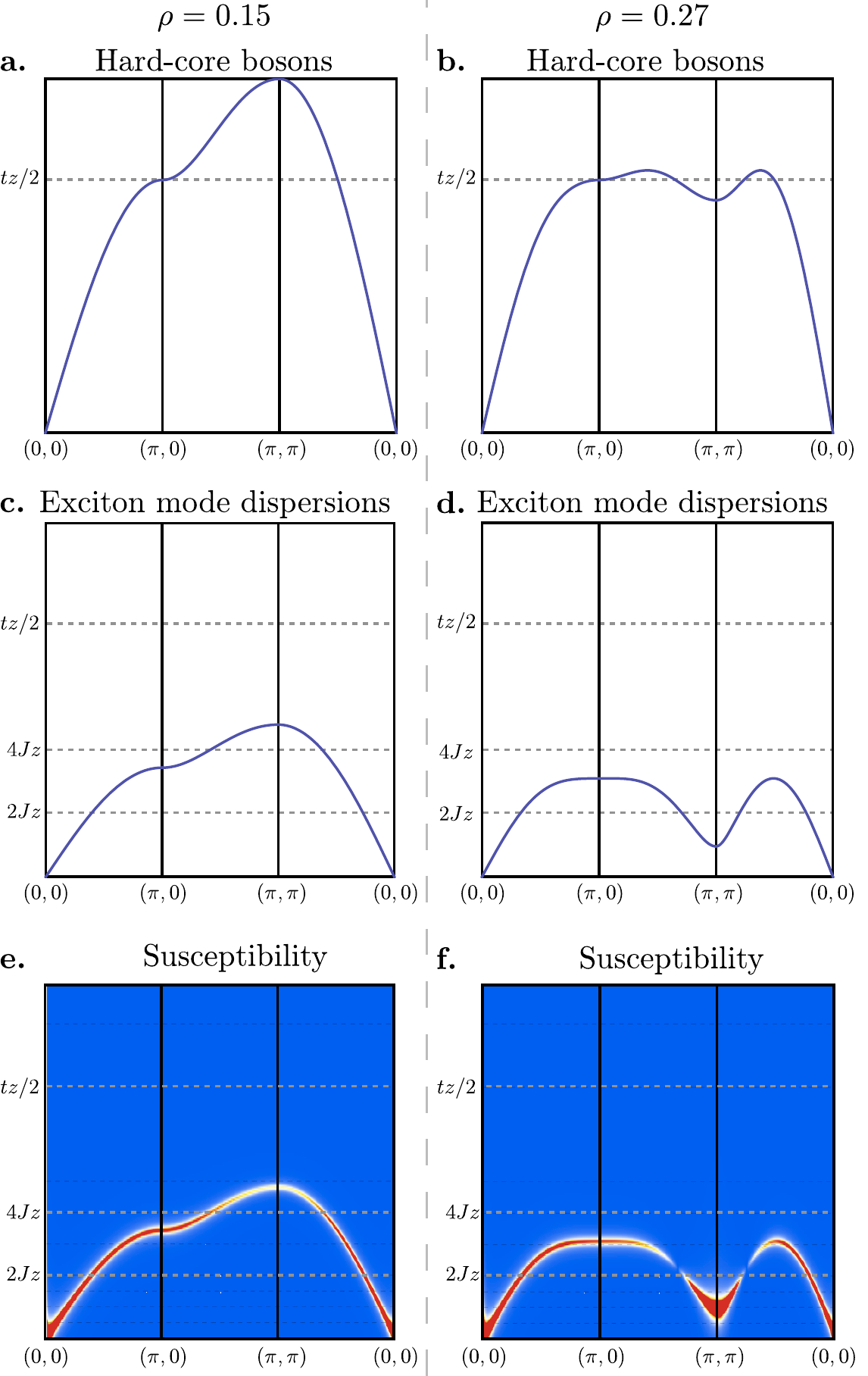}
	\caption{\label{EC-exciton}
	Dispersions and susceptibilities of the Goldstone mode associated with the exciton condensate. We have set $t=V=2$ eV, $J=125$ meV and $\alpha=0.04$, and the exciton density is either $\rho=0.15$ (left column) or $\rho=0.27$ (right column). \textbf{a, b.} In the simple hard-core boson model the condensate phase clearly show the superfluid phase mode, linear at small momenta. \textbf{c,d.} In the full $t-J$ model the Goldstone mode has a similar dispersion as in the $XXZ$ model. The speed of the mode scales with the superfluid density. At higher densities the mode softens around $(\pi,\pi)$, and when this gap closes a first order transition to the checkerboard phase sets in. \textbf{e,f.} The absorptive part of the charge susceptibility, which can be measured with for example EELS or RIXS.}
\end{figure}

\begin{figure}
	\includegraphics[width=\columnwidth]{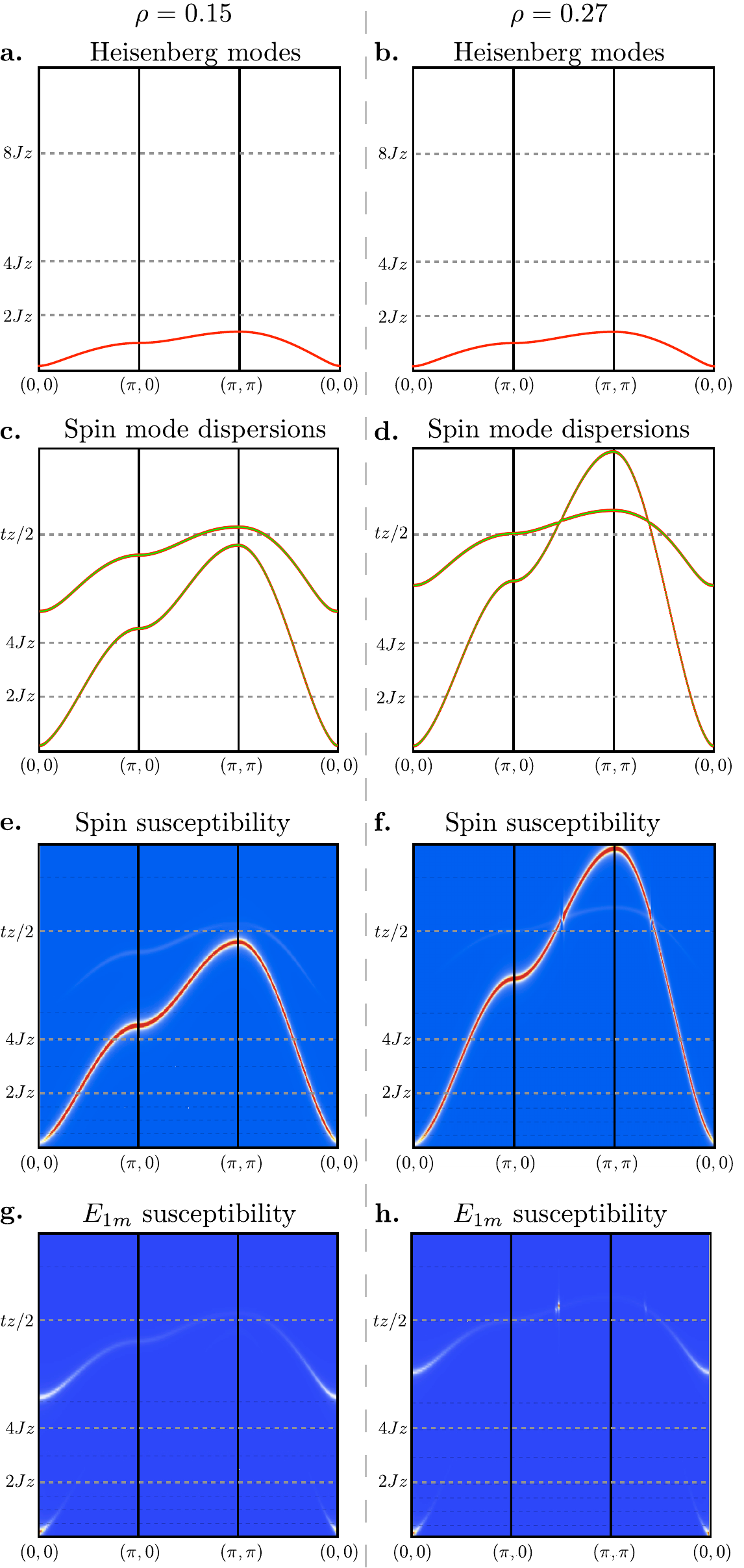}
	\caption{\label{EC-spin} 
	Dispersions and magnetic susceptibilities of the exciton condensate. We have set $t=V=2$ eV, $J=125$ meV and $\alpha=0.04$, and the exciton density is either $\rho=0.15$ (left column) or $\rho=0.27$ (right column).
	\textbf{a,b.} As the exciton condensate is spin singlet, we assume that the excitation spectrum is governed by propagating triplet modes. These modes have a gap of order $J_\perp$ and a bandwidth of order $Jz$.
	\textbf{c,d.} In contrast to the simple Heisenberg results, the actual triplet modes have enhanced kinetics\cite{RademakerMarch2013}. The modes are split in a spin-dominated branch with small gap and large bandwidth proportional to the superfluid density (\textbf{e,f.}); and an exciton-dominated branch with a large gap and a small bandwidth  (\textbf{g,h.}).}
\end{figure}

The mode spectrum of superfluid phase, as shown in figures \ref{EC-exciton} and \ref{EC-spin}, is characterized by a linearly dispersing Goldstone mode associated with the broken $U(1)$ symmetry. This superfluid phase mode has vanishing energy at the $\Gamma$ point, where we find the inescapable linear dispersion relation
\begin{equation}
	\omega_k = \frac{1}{4\sqrt{2}} zt \sqrt{(1-\rho)\rho \; (1+2V/t)} \; |k| + \ldots
\end{equation}
The speed of the superfluid phase mode is the same as for the $XXZ$ model in equation (\ref{XXZsuperfluidSpeed}) up to a rescaling of the $t$ and $V$ parameters. Indeed, this speed is proportional to the superfluid density $\sqrt{\rho_{\mathrm{SF}}} = \sqrt{\rho(1-\rho)}$. This mode can be seen in the charge susceptibility, figures \ref{EC-exciton}e and f. The Goldstone mode has a gap at $(\pi,\pi)$ which decreases monotonically with increasing exciton density. Precisely at the first order transition towards the checkerboard phase this gap closes. This mode softening at $(\pi,\pi)$ is reminiscent of the roton in superfluid Helium: the wavelength of the roton is the same as the lattice constant of solid Helium.

Next to the Goldstone mode there are two triplet excitations, shown in figure \ref{EC-spin}, each one three-fold degenerate. The degeneracy obviously arises from the standard triplet degeneracy $m=-1, 0, +1$. The two branches however distinguish between \emph{exciton-dominated} modes and \emph{spin-dominated} modes, let us discuss them separately.

The spin-dominated modes have a gap of order $\Delta_S = Jz \sqrt{\alpha(1+\alpha-\rho)}$, which is similar to the triplet gap in the bilayer Heisenberg model for large $\alpha$. However, the bandwidth of these excitations scales with $t$ rather than with $J$, as would be customary in a system without exciton condensation (see figures \ref{EC-spin}a and b). We discussed this in great detail in recent work\cite{RademakerMarch2013}, so let us briefly review these results. In the absence of a excitons the motion of triplets is governed by the Heisenberg superexchange yielding a bandwidth of order $J$. Now introduce Fock operators $e^\dagger = |E \rangle \langle 0 |$ and $t^\dagger = |1m \rangle \langle 0 |$, so that the exciton-triplet exchange equation (\ref{ExcitonHop}) reads
\begin{equation}
	-t \sum_{\langle ij \rangle} e_j^\dagger e_i t^\dagger_i t_j.
\end{equation}
This is an interaction term, thus seemingly irrelevant to the bandwidth of the triplet. However, when the exciton condensation sets in the operator $e^\dagger$ obtains an expectation value, in fact $\langle e^\dagger \rangle = \sqrt{\rho_{\mathrm{SF}}}$ where $\rho_{\mathrm{SF}}$ is the condensate density. Therefore the higher order exchange term yields a quadratic triplet hopping term
\begin{equation}
	-t \rho_{\mathrm{SF}} \sum_{\langle ij \rangle} t^\dagger_i t_j
\end{equation}
and the bandwidth of the triplet excitations becomes of order $zt \rho_{\mathrm{SF}}$.
Now remember that the exciton hopping energy $t$ resulted, in second order perturbation theory, from the ratio $t_e^2/V'$ where $t_e$ is the electron hopping energy and $V'$ is the nearest neighbor Coulomb repulsion\cite{Rademaker2012EPL,Rademaker2012arXiv}. The Heisenberg coupling however was given by $J=2t^2/U$ where $U$ is the onsite Coulomb repulsion. Since for obvious reasons $U > V'$, we find that the triplet bandwidth is enhanced whenever exciton condensation sets is. This enhancement is clearly visible in the spin susceptibility $\chi_{T}''$, which allows for an experimental probe of the exciton superfluid density.

The other branch of triplet excitations is dominated by triplet excitons, and is therefore barely visible in the spin susceptibility and not visible in the exciton susceptibility (which only shows singlet excitons). That it is indeed dominated by triplet excitons can be inferred from computing the matrix elements of the operator $E_{1m}$, which are shown in figures \ref{EC-spin}g and h. Furthermore, the gap $\Delta_E = (Vz+tz)\rho - \mu$ is a function of exciton model parameters only. The bandwidth of this mode is of order $\mathcal{O}(zt)$, relatively independent of the exciton density. As a result, for large superfluid densities the exciton-dominated modes cross the spin-dominated triplet modes. One can directly see this in the excitation spectrum for $\rho=0.27$ as shown in figure \ref{EC-spin}d.

We can compare the triplet spectrum to the mode spectrum of the singlet phase of the bilayer Heisenberg model. When $J_\perp \gg J$ the ground state consists of only rung singlets. The excitation towards a triplet state, shown in figures \ref{EC-spin}a and b, has a gap $Jz \sqrt{\alpha (\alpha - 1)}$ and a bandwidth of order $Jz$, which is considerably smaller than the $\mathcal{O}(zt)$ bandwidth in the condensate. However, because the topology of the triplet mode is the same we expect that the effect of the spin-exciton interactions is the same in the bilayer Heisenberg model as for the superfluid. Since earlier LSW-SCBA showed no changes in the spectrum due to interactions, we infer that the non-interacting results for the superfluid are reliable. 

To conclude our discussion of the excitations of the superfluid phase let us consider the influence of the interlayer tunneling. In the context of the $XXZ$ model we noticed that interlayer tunneling has no qualitative influence on the phase diagram itself. However, the presence of a weak interlayer tunneling may act as potential pinningthe phase\cite{RademakerPRB2011} opening a gap in the superfluid phase mode spectrum of order $\mathcal{O} (\sqrt{t_\perp (V+t)})$. Persistent currents can still exist, but one needs to overcome this gap in order to get the exciton supercurrent flowing.

\subsection{Checkerboard phase}

\begin{figure}
	\includegraphics[width=\columnwidth]{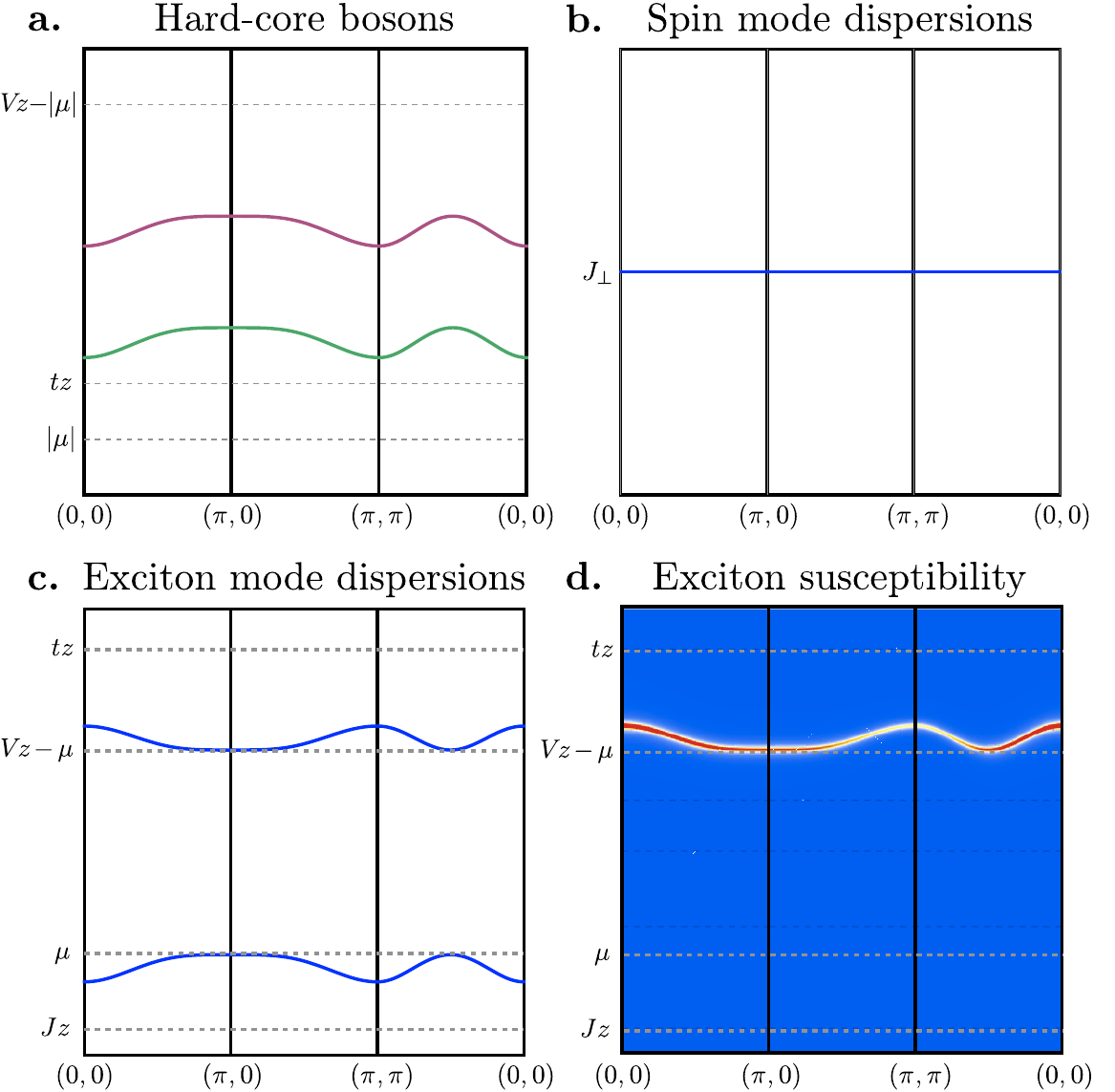}
	\caption{\label{DispersionFigCB}
	The excitation spectrum of the checkerboard phase. \textbf{a.} In the simple hard-core boson model there are two exciton modes associated with the 'doublon' and the 'holon' excitation. \textbf{b.} The spin modes are decoupled from the exciton modes in the full $t-J$ model. There is only one possible spin excitation: changing the singlet groundstate into a non-propagating triplet. \textbf{c.} The exciton modes, on the other hand, can still propagate. The excitation of removing an exciton can propagate through the checkerboard. \textbf{d.} The propagating mode that changes an exciton into a singlet is detectable by optical means and thus shows up in the charge susceptibility.}
\end{figure}

The third homogeneous phase of the exciton $t-J$ model is the checkerboard phase. In this phase the unit cell is effectively doubled with one exciton per unit cell. This state is analogous to a Bose Mott insulator. The trivial excitations are then the doublon and the holon: create two bosons per unit cell which costs an energy $Vz-\mu$ or to remove the boson. The latter will generate a propagating exciton mode, with dispersion
\begin{equation}
	\omega_{k,pm} = 
		\frac{1}{2} \left( \pm Vz 
			+ \sqrt{ (Vz)^2 \pm \left(\frac{1}{2} z t \gamma_k \right)^2} \right)
			\mp \mu \pm J_\perp.
\end{equation}
There are two such propagating modes: one associated with the singlet exciton and one with the triplet exciton. Precisely at the transition towards the superfluid phase, one of these exciton waves becomes gapless. Note that the arguments that lead to the bandwidth renormalization in the antiferromagnetic phase also apply here, leading to an exciton bandwidth of order $t^2/V$. The dispersions and the corresponding charge dynamical susceptibility can be seen in figure \ref{DispersionFigCB}.

In the spin sector one can excite a localized spin triplet on the empty sublattice. The triplet gap is set by the interlayer energy $J_\perp$, and the dispersion is flat because this triplet cannot propagate, as can be seen in figure \ref{DispersionFigCB}b.

\section{Conclusions and discussion}
\label{SecConc}

We have studied the possibility of exciton condensation in strongly correlated electron bilayers. Starting from the description of the Mott state, with localized electrons, an exciton is defined as an interlayer bound state of a double occupied and vacant site. In the strong coupling limit, as of relevance to laboratory systems based on Mott insulators, the physics of such a system is described by the exciton $t-J$ model, equation (\ref{FullH}). 

We constructed the ground state phase diagram (figure \ref{FinalPhaseDiagramFig}), based on both numerical simulations and analytical mean field theory. Three distinct phases are dominant: the antiferromagnetic phase, the checkerboard phase and the exciton condensate. For most parts of the phase diagram however, macroscopic phase separation will occur between these three phases. 

Measurements of the spin and charge susceptibilities may discern in which one of the three main phases a specific system is in. The antiferromagnetic phase is characterized by a spin wave centered at $(\pi,\pi)$, whilst in the exciton condensate the triplet bandwidth acts as a probe for the superfluid density (see figure \ref{EC-spin} and Ref. \cite{RademakerMarch2013}). In the checkerboard phase the spin degrees of freedom are reflected only in a localized triplet excitation at low energy.

The charge dynamic susceptibility shows distinct qualitative behavior depending on the phase. In the antiferromagnet the spin-exciton interactions play an important role \cite{Rademaker2012EPL, Rademaker2012arXiv}. The superfluid phase is characterized by the visibility of the condensate Goldstone mode, whereas the checkerboard phase has propagating exciton waves with bandwidth $\mathcal{O}(t^2/V)$. Note however that since we expect phase separation to occur for most model parameters, realistic samples will likely display features from all phases in its susceptibilities.

Our theoretical work presented here is largely based on the assumption of strong coupling. In this limit, the excitons behave as local hard-core bosons. If the exciton binding energy is less dominant, the exciton will extend over more lattice sites and thus probably enable coexistence phases. On the other hand we expect that spin-exciton interactions destabilize the coexistence phases, since these interactions generally lead to frustration effects. One could also wonder what happens if one includes longer-ranged interactions for the excitons, with the possibility of exciton stripes and incommensurate charge ordered phases\cite{Rademaker:2013p5682}. Next, we are dealing with first order phase transitions where small changes may have severe consequences. Combining all these effects may lead to significant changes in the phase diagram, most notably in the regime where we predict phase separation.

Within the context of the strongly coupled exciton $t-J$ model, a weaker exciton binding energy can be incorporated via interaction and hopping terms for the next nearest neighbors, next next nearest neighbors, etcetera. This might lead to complex ordered phases such as stripes\cite{Parish:2012p5404}. Such phases are found in many strongly correlated electron systems\cite{Zaanen:1989p1602,Tranquada:1995p5405}, and studying these in the context of the simple bosonic exciton $t-J$ model might shed new light on the more troublesome fermionic $t-J$ model.

In addition to stripy behavior other non-trivial exciton density profiles may occur when one considers a density imbalance between the electrons and holes. Semiconductor imbalanced systems are predicted to exhibit Fulde-Ferrell-Larkin-Ovchinnikov density modulated phases\cite{Subasi:2010p4927,Parish:2011p5107}. It is worthwhile to investigate whether such phase can exist in strongly correlated electron systems.

Next to an improvement of the phase diagram, we can also improve the susceptibilities by including the effect of exciton-spin interactions. Similar to our earlier work \cite{Rademaker2012EPL,Rademaker2012arXiv} on the interaction between excitons and spins in the limit of a single exciton, one could perform a diagrammatic expansion of these interactions. We expect that, apart from our earlier results in the antiferromagnetic phase, inclusion of spin-exciton interactions will not qualitatively alter the excitation spectra.

Experimentally, the close coupling of p- and n-doped Mott insulators is still relatively ill explored. However, important advances in complex oxide thin film growth, by techniques such as Molecular Beam Epitaxy (MBE) and Pulsed Laser Deposition (PLD) equipped with in situ monitoring tools such as Reflective High Energy Electron Diffraction (RHEED) are making it possible now to grow multilayers of perovskite oxides - of which many are Mott insulators - with unit cell precision. A complicating factor in fabricating multilayers of p- and n-doped perovskites, like the cuprate family from which also the high-Tc superconductors are derived, are the oftentimes conflicting (de)-oxygenation requirements. Optimized deposition and post-anneal procedures have made it possible however to make thin film contacts between n- and p-doped superconducting cuprates \cite{Takeuchi1995}, which is now further being explored in our labs to create and study the parallel n-p combinations resembling the theoretical model.

An interesting additional system that can be included in this endeavor is the 2-dimensional electron gas that is formed at the interfaces between selected oxide band-insulators such as SrTiO$_3$ and LaAlO$_3$. In this respect it is noteworthy that in specific configurations, in particular a 1 unit cell SrTiO$_3$ layer on top of a 2 unit cell LaAlO$_3$ layer grown on TiO$_2$-terminated SrTiO$-3$, a system of a closely coupled 2-dimensional electron gas and a 2-dimensional hole gas has been realized \cite{Pentcheva:2009p5025}. 

Finally we note that some cuprate high-Tc materials appear to have an intrinsic stacking of electron-doped and hole-doped CuO$_2$ layers, such as Ba$_2$Ca$_3$Cu$_4$O$_8$F$_2$ \cite{Chen2006}, where one could look for excitonic effects.

\acknowledgments

This research was supported by the Dutch NWO foundation through a VICI grant. The authors thank Sergei Mukhin and Kai Wu for helpful discussions.

\appendix

\section{$SU(5)$ structure of the exciton $t-J$ model}
\label{AppendixA}
In this appendix we will define the operators that compose the $SU(5)$ dynamical Lie algebra, as described in the beginning of section \ref{AlgebraSec}. From the bilayer Heisenberg we already have the $SO(4)$ spin subalgebra
\begin{eqnarray}
	S^z & = & | 1 + \rangle \langle 1 + |- | 1 - \rangle \langle 1 - | \\
	S^+ & = & \sqrt{2} \left( | 1 + \rangle \langle 1 0 | 
		+ | 1 0 \rangle \langle 1 - | \right)\\
	\widetilde{S}^z & = & - | 0 \rangle \langle 1 0 |- | 10 \rangle \langle 0 | \\
	\widetilde{S}^+ & = & \sqrt{2} \left( | 1 + \rangle \langle 0 | 
		- |  0 \rangle \langle 1 - | \right)
\end{eqnarray}
where we use the obvious short-handed notation for the singlet and triplet kets and bras. The commutation relation between these operators read
\begin{eqnarray}
	\left[ S^a, S^b \right] &=& i \epsilon^{abc} S^c \\
	\left[ S^a, \widetilde{S}^b \right] &=& \left[ \widetilde{S}^a, S^b \right] 
		= i \epsilon^{abc} \widetilde{S}^c \\
	\left[ \widetilde{S}^a, \widetilde{S}^b \right] &=& i \epsilon^{abc} S^c.
\end{eqnarray}
There are 12 exciton operators in the $XY$-like part of the Hamiltonian, which we denote by 
\begin{eqnarray}
	E^+_{sm} & = & | E \rangle \langle s \; m | \\
	E^z_{sm} & = & \frac{1}{2} \left( 
		| E \rangle \langle E | - | s \; m \rangle \langle s \; m | \right),
\end{eqnarray}
where $s=0,1$ and $m=-s \ldots +s$, with commutation relations
\begin{eqnarray}
	\left[ E^+_{sm}, E^-_{sm} \right] &=& 2E^z_{sm} \\
	\left[ E^z_{sm},E^+_{sm} \right] &=& E^+_{sm}.
\end{eqnarray}
If we consider commutators between $E$-operators with different $sm$ then we obtain operators that are fully spin-dependent. We find that the only nonzero commutators between $E_{sm}$ operators with $ms \neq m's'$,
\begin{eqnarray}
	\left[ E^-_{sm}, E^+_{s'm'} \right] &=& | s \; m \rangle \langle s' \; m' | \\
	\left[ E^z_{sm},E^+_{s'm'} \right] &=&\frac{1}{2} E^+_{s'm'}.
\end{eqnarray}
We need some more operators to close the spin $SU(4)$ subalgebra, therefore define $T, \widetilde{T}$ operators by (note that $(\widetilde{T}^z)^\dagger = -\widetilde{T}^z$),
\begin{eqnarray}
	T^+ & = & \sqrt{2} \left( | 1 + \rangle \langle 1 0 | 
		- |  1 0 \rangle \langle 1 - | \right) \\
	\widetilde{T}^z & = &  | 0 \rangle \langle 1 0 |- | 10 \rangle \langle 0 | \\
	\widetilde{T}^+ & = & \sqrt{2} \left( | 1 + \rangle \langle 0 | 
		+ |  0 \rangle \langle 1 - | \right).
\end{eqnarray}
To complete the spin $SU(4)$ subalgebra we define
\begin{eqnarray}
	M^+ & = & | 1 + \rangle \langle 1 - |
\end{eqnarray}
and the corresponding $M^- = (M^+)^\dagger$. Finally, notice that we have one operator too much in our listing, since there are only 24 operators in $SU(5)$. Thus there exists a linear dependency relation between some operators, which is
\begin{equation}
	\frac{1}{2} S^z + E^z_{1+} - E^z_{1-} = 0.
\end{equation}
So when constructing the Heisenberg equations of motion, we will exclude one of these three from our formalism. The most logical step is to throw out the combination $E^z_{1+} - E^z_{1-}$ and leave the sum
\begin{equation}
	E^z_{1m} \equiv E^z_{1+} + E^z_{1-}.
\end{equation}
The 24 operators of our dynamical $SU(5)$ algebra are the three $S$-, three $\widetilde{S}$-, two $T$-, three $\widetilde{T}$-, two $M$- and eleven $E$-operators.

\section{The Heisenberg equations of motion method}
\label{AppendixC}

In this appendix we elaborate a bit further on the Heisenberg equations of motion method, as introduced in paragraph \ref{XXZexc}. The aim of this method is to find the spectrum of excitations, building on the foundations given by the mean field approximation. Given a full set of local operators $\mathcal{A}^{\ell}_i$, we can construct the Heisenberg equations of motion
\begin{equation}
	i \partial_t \mathcal{A}^{\ell}_i = [\mathcal{A}^{\ell}_i, H]
\end{equation}
which is in general impossible to solve. We employ the notation that $i$ indicates the lattice site, and $\ell$ is the index denoting the type of operator. The right hand side of this equation contains products of operators at different lattice sites. Such products can be decoupled within the mean field approximation as
\cite{Zubarev1960,Oles:2000p5088}
\begin{equation}
	\mathcal{A}^{\ell}_i \mathcal{A}^{\ell'}_j
		\rightarrow \langle \mathcal{A}^{\ell}_i \rangle \mathcal{A}^{\ell'}_j 
		+ \mathcal{A}^{\ell}_i \langle \mathcal{A}^{\ell'}_j \rangle
\end{equation}
where $i$ and $j$ are different lattice sites. Upon Fourier transforming lattice position into momentum and time into energy, we thus obtain a set of linear equations for the operators,
\begin{equation}
	\omega_q \mathcal{A}^{\ell} (q, \omega) = M^{\ell \ell'} (q) \mathcal{A}^{\ell'} (q, \omega).
\end{equation}
The spectrum of excitations is simply found by solving this eigenvalue equation for the matrix $M(q)$.

In order to find the matrix elements $\langle n | \mathcal{A}^{\ell} (q) | 0 \rangle$ that enter in susceptibilities we need will introduce the following scheme. Assume that the Hamiltonian is of the form
\begin{equation}
	H = \sum_{qn} \omega_{qn} \alpha^\dagger_{qn} \alpha_{qn}
\end{equation}
where the sum over $q$ runs over momenta, and $n$ indicates the different excited states. Now $\alpha^\dagger_{qn}$ is a creation operator, and irrespective of whether we are dealing with fermions or bosons we have the following equations of motion
\begin{equation}
	i \partial_t \alpha^\dagger_{qn} = - \omega_{qn} \alpha^\dagger_{qn}.
\end{equation}
That is: every eigenvector of $M^{\ell \ell'} (q)$ corresponding to a negative eigenvalue can be identified as a creation operator for one of the elementary excitations. However, the eigenvalue equation itself is not enough because it does not yield the proper normalization of $\alpha^\dagger$. Since we have the eigenvector solution
\begin{equation}
	\alpha^\dagger_{qn} = \mathcal{U}^{n \ell} \mathcal{A}^{\ell} (q)
\end{equation}
we can write out the (anti)commutation relation for $\alpha^\dagger_{qn}$ in terms of the (anti)commutation relations for the $\mathcal{A}^{\ell} (q)$. Upon requiring that on the mean field level the operators $\alpha^\dagger_{qn}$ obey canonical commutation relations, that is for bosons
\begin{equation}
	\langle \left[ \alpha_{qn}, \alpha^\dagger_{qn'} \right] \rangle = \delta_{nn'},
\end{equation}
we obtain a proper normalization for the new creation operators. We can invert the normalized matrix $\mathcal{U}^{n \ell}$ to express $\mathcal{A}^{\ell} (q)$ in terms of the creation operators $\alpha^\dagger_{qn}$. Finally, using $\langle n' | \alpha^\dagger_{qn} | 0 \rangle = \delta_{nn'}$ we can compute the wanted matrix element for $\mathcal{A}^{\ell} (q)$.

As an example of this technique we can compute the matrix element $\left| \langle n | S^+(q)|0\rangle \right|^2$ for the antiferromagnetic Heisenberg model on a square lattice. The mean field ground state is the N\'{e}el state, which leads to the following equations of motion,
\begin{equation}
	 i \partial_t \begin{pmatrix} S^+_{qA} \\ S^+_{qB} \end{pmatrix}
	 	= \frac{1}{2} J z
		\begin{pmatrix} 1 & \gamma_q \\ -\gamma_q & -1 \end{pmatrix}
		\begin{pmatrix} S^+_{qA} \\ S^+_{qB} \end{pmatrix}.
\end{equation}
where the subscript $A$ and $B$ denote the two different sublattices, and $\gamma_q = \frac{1}{2} ( \cos q_x + \cos q_y )$. We quite easily infer that the eigenvalues are
\begin{equation}
	\omega_q = \pm \frac{1}{2} Jz \sqrt{ 1- \gamma_{q}^2}
\end{equation}
and thus we have one eigenvector corresponding to a creation operator, and one to an annihilation operator. If we define
\begin{equation}
	\begin{pmatrix} \alpha^\dagger \\ \beta \end{pmatrix}
		= \mathcal{U} \begin{pmatrix} S^+_{qA} \\ S^+_{qB} \end{pmatrix}
\end{equation}
then the commutation relations tell us that the eigenvector matrix $\mathcal{U}$ must satisfy
\begin{equation}
	1 = \langle [ \alpha, \alpha^\dagger] \rangle
	= - 2 u_{11}^2 \langle S^z_{A} \rangle - 2 u_{12}^2 \langle S^z_{B} \rangle 
	= - u_{11}^2 + u_{12}^2.
\end{equation}
The initial $S^+_q$ operator, which enters in the spin susceptibility, can be expressed in terms of the eigenvector matrix as
\begin{equation}
	\mathcal{S}^+_q = \frac{1}{\sqrt{2}} ( 1 \; 1) \; \mathcal{U}^{-1} 
		\begin{pmatrix} \alpha^\dagger \\ \beta \end{pmatrix}.
\end{equation}
Some straightforward algebra now yields 
\begin{equation}
	\left| \langle n | S^+(q)|0\rangle \right|^2
		= \frac{1}{2} \sqrt{\frac{1-\gamma_q}{1+\gamma_q}}
\end{equation}
which is the same susceptibility one can obtain by using the Holstein-Primakoff linear spin wave approximation. The approximation scheme we introduced here can therefore be viewed as a generalization of the linear spin wave approximation.

\end{document}